\documentclass[acmsmall]{acmart}

\usepackage{algorithm}
\usepackage{algpseudocode}
\usepackage{caption}
\usepackage{glossaries}
\usepackage{setspace}
\usepackage{multirow}
\usepackage{makecell}
\usepackage{xspace}
\usepackage{appendix}
\usepackage{soul}
\usepackage{enumitem}
\usepackage{graphicx}
\usepackage{subcaption}
\usepackage{array}
\usepackage{ulem}
\usepackage{wrapfig}
\usepackage{amsmath}

\newacronym{GRU}{GRU}{Gated Recurrent Unit}
\newacronym{A-DDPG}{A-DDPG}{Adversarial Deep Deterministic Policy Gradient}
\newacronym{DDPG}{DDPG}{Deep Deterministic Policy Gradient}
\newacronym{MSE}{MSE}{Mean Squared Error}
\newacronym{MLP}{MLP}{Multi-Layer Perceptron}
\newacronym{DT}{DT}{Decision Tree}
\newacronym{RF}{RF}{Random Forest}
\newacronym{LSTM}{LSTM}{Long Short-Term Memory}
\newacronym{Bi-GRU}{Bi-GRU}{Bidirectional Gated Recurrent Unit}
\newacronym{CNN}{CNN}{Convolutional Neural Network}
\newacronym{SVM}{SVM}{Support Vector Machine}
\newacronym{GNB}{GNB}{Gaussian-kernel Naive Bayes}
\newacronym{RL}{RL}{Reinforcement Learning}
\newacronym{TDE}{TDE}{Temporal Difference Error}
\newacronym{NMAE}{NMAE}{Noromalized Mean Absolute Errors}
\newacronym{ASR}{ASR}{Attack Success Rate}
\newacronym{ECDF}{ECDF}{Empirical Cumulative Distribution Function}
\newacronym{DPI}{DPI}{Deep Packet Inspection}
\newacronym{GFW}{GFW}{Great Firewall}
\newacronym{CDN}{CDN}{Content Delivery Networks}
\newacronym{GAN}{GAN}{Generative Adversarial Networks}
\newacronym{ML}{ML}{Machine Learning}
\newacronym{FGSM}{FGSM}{Fast Gradient Sign Method}
\newacronym{NIDS}{NIDS}{Network Intrusion Detection Systems}
\newacronym{DQN}{DQN}{Deep Q Networks}
\newacronym{PSA}{PSA}{Proportional Scaling Attack}
\newacronym{NN}{NN}{Neural Network}
\newacronym{TCBs}{TCBs}{Transmission Control Blocks}
\newacronym{CW}{CW}{Carlini \& Wagner}
\newacronym{PPO}{PPO}{Proximal Policy Optimization}
\newacronym{TRPO}{TRPO}{Trust Region Policy Optimization}
\newacronym{GAE}{GAE}{Generalized Advantage Estimation}
\newacronym{DF}{DF}{Deep Fingerprinting}
\newacronym{SDAE}{SDAE}{Stacked Denoising Autoencoder}
\newacronym{WF}{WF}{Website Fingerprinting}
\newacronym{BAP}{BAP}{Blind Adversarial Perturbation}

\colorlet{RED}{red}

\newcolumntype{?}{!{\vrule width 1pt}}
\algnewcommand{\Inputs}[1]{%
  \State \textbf{Inputs:}
  \Statex \hspace*{\algorithmicindent}\parbox[t]{.8\linewidth}{\raggedright #1}\vspace{3pt}
}
\algnewcommand{\Initialize}[1]{%
  \State \textbf{Initialisation:}
  \Statex \hspace*{\algorithmicindent}\parbox[t]{.8\linewidth}{\raggedright #1}\vspace{3pt}
}

\AtBeginDocument{%
  \providecommand\BibTeX{{%
    \normalfont B\kern-0.5em{\scshape i\kern-0.25em b}\kern-0.8em\TeX}}}

\begin{document}

\title{Amoeba: Circumventing ML-supported Network Censorship via Adversarial Reinforcement Learning}

\author{Haoyu Liu}
\email{haoyu.liu@ed.ac.uk}
\affiliation{%
  \institution{The University of Edinburgh}
  \streetaddress{2 Charles St}
  \city{Edinburgh}
  \country{United Kingdom}
  \postcode{EH8 9AD}
}

\author{Alec F. Diallo}
\email{alec.frenn@ed.ac.uk}
\affiliation{%
  \institution{The University of Edinburgh}
  \streetaddress{2 Charles St}
  \city{Edinburgh}
  \country{United Kingdom}
  \postcode{EH8 9AD}
}

\author{Paul Patras}
\email{paul.patras@ed.ac.uk}
\affiliation{%
  \institution{The University of Edinburgh}
  \streetaddress{2 Charles St}
  \city{Edinburgh}
  \country{United Kingdom}
  \postcode{EH8 9AD}
}
\renewcommand{\shortauthors}{Haoyu Liu, Alec F. Diallo, \& Paul Patras}

\begin{abstract}
Embedding covert streams into a cover channel is a common approach to circumventing Internet censorship, due to censors' inability to examine encrypted information in otherwise permitted protocols (Skype, HTTPS, etc.). However, recent advances in machine learning (ML) enable detectin g a range of anti-censorship systems by learning distinct statistical patterns hidden in traffic flows. Therefore, designing obfuscation solutions able to generate traffic that is statistically similar to innocuous network activity, in order to deceive ML-based classifiers at line speed, is difficult.

In this paper, we formulate a practical adversarial attack strategy against flow classifiers as a method for circumventing censorship. Specifically, we cast the problem of finding adversarial flows that will be misclassified as a sequence generation task, which we solve with Amoeba, a novel reinforcement learning algorithm that we design. Amoeba works by interacting with censoring classifiers \textit{without any knowledge of their model structure}, but by crafting packets and observing the classifiers' decisions, in order to guide the sequence generation process. Our experiments using data collected from two popular anti-censorship systems demonstrate that Amoeba can effectively shape adversarial flows that have on average 94\% attack success rate against a range of ML algorithms. In addition, we show that these adversarial flows {are robust in different network environments} and possess transferability across various ML models, meaning that once trained against one, our agent can subvert other censoring classifiers without retraining. 
\vspace{-0.5ex}
\end{abstract}


\setcopyright{acmlicensed}
\acmJournal{PACMNET}
\acmYear{2023} \acmVolume{1} \acmNumber{3} \acmArticle{9} \acmMonth{12} \acmPrice{15.00}\acmDOI{10.1145/3629131}


\settopmatter{printfolios=true}
\maketitle
\section{Introduction}
Governments and control authorities in some countries carry out network traffic censorship routinely to restrict the citizens' access to online information that may be perceived by those regimes as politically, socially, or morally objectable. To maintain censorship effectiveness and combat circumvention, e.g., through traffic mimicry and randomization~\cite{wang2015seeing}, state actors employ a range of tools including \gls{DPI}, protocol fingerprinting, and active probing. In recent years, protocol tunneling has gained traction as a viable means to circumvent censorship. Tunneling leverages existing implementations of innocuous protocols (Skype, WebRTC, TLS, etc.) and embeds covert streams in these protocols to hide destination host identity, payload contents, etc. \cite{barradas2017deltashaper, barradas2020poking, frolov2019use}. As a result, sensitive information becomes encrypted and message exchanges perfectly aligned with the behavior of the tunneling protocol. In turn, observing the tunnels barely unveils any deterministic fingerprints. 
Censorship is however an arms race, recent studies revealing that \gls{ML} algorithms, which learn statistical features from network flows, can effectively identify `offending' tunneled traffic, despite not exhibiting deterministic fingerprints \cite{deng2017random, wang2015seeing}. 
For example, although multiple multi-media tunneling tools claim unobservability, simple ML classifiers such as those based on decision tree and random forest structures can detect tunneled traffic with high confidence \cite{barradas2018effective}. \textcolor{black}{On the other hand, \gls{ML} is also employed to devise censorship circumvention strategies. For example, Geneva \cite{bock2019geneva} designs a genetic algorithm to 
discover if existing censorship can be evaded by tampering with canonical TCP implementations, e.g., by corrupting checksums, breaking \gls{TCBs} (by injecting a \texttt{RST}), or segmenting packets with corrupted \texttt{ACK}s. While this attack targets the incompleteness of network stacks implemented by censors, in this paper we aim to push the boundary of anti-censorship one step further, where we reasonably assume the censor fixes the implementation issues and leverages ML classifiers to detect anti-censorship tools.}

\begin{figure*}[t]
    \centering
    \includegraphics[width=0.9\linewidth]{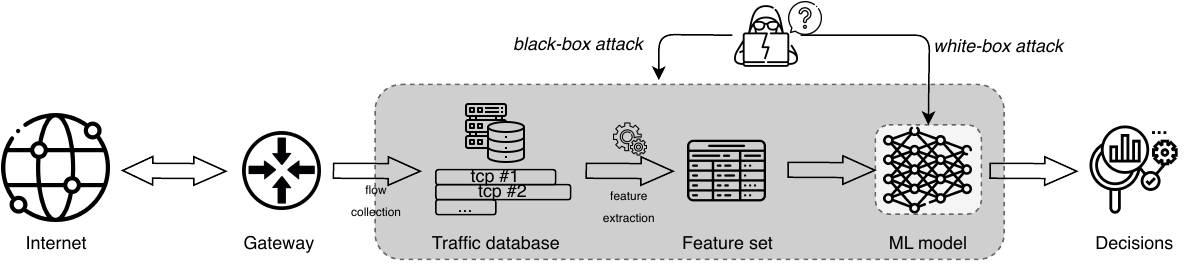}
    \vspace*{-1em}
    \caption{Typical traffic classification pipeline. An attacker may conduct adversarial attacks against the pipeline with different capabilities/scope. We categorize them into 1) \textit{white-box model} for which the inner workings of the classifier (weights, gradients) are visible; and 2) \textit{black-box model} for which attackers can only craft network flows and observe outputs without extra knowledge, such as feature engineering and model architecture.}
    \label{fig:adv_threat_models}
    \vspace*{-2em}
\end{figure*}

Since the inner workings of an ML-based classifier are largely unknown to users seeking to circumvent censorship and the censor can change the underlying neural architecture at any time (black box), the question we aim to answer in this work is: \textit{Instead of perpetually designing new tunneling tools, can we devise adversarial attacks on {black-box} ML classifiers to consistently subvert ML-supported censorship?} 
This approach has not been well studied in the network censorship domain. In computer vision, finding adversarial examples, i.e., images that should be recognized as belonging to class A being instead misclassified as of class B, can be achieved by adding adversarial perturbations such that the modified input images remain visually similar to their original versions, but produce erroneous classification results \cite{goodfellow2014explaining,andriushchenko2020square,carlini2017towards}.

Conducting adversarial attacks in the networking domain is fundamentally different. A common approach to ML-based network flow classification is to first extract multiple statistical features (packet size distributions, timing information, etc.), then feed these features to a classifier instead of raw flows \cite{liu2022netsentry, barradas2018effective}, as illustrated in Figure \ref{fig:adv_threat_models}. Censors do not reveal what features they utilize, which poses difficulty in the first step of crafting an adversarial attack. Further, even if users may discover the set of features employed by a censoring classifier and successfully generate adversarial samples, there is no guarantee that such samples can be mapped back to legitimate flows, which renders the entire process unusable in practice. A practical adversarial attack against censoring classifiers requires manipulation at packet level, instead of feature level, and each packet should be transmitted without adding significant delays. Early attempts \cite{verma2018network,nasr2021defeating} apply attacks on complete flows and generate adversarial versions, yet each manipulated packet should be sent before new packets are received. Having a complete view of a flow to perturb is unfortunately unrealistic. The inherent imbalance between what censors can observe (flows) and what users can observe and manipulate (packets) rules out the possibility of applying existing algorithms from other domains to achieve adversarial attacks for censorship circumvention purposes. 

In this paper, we formulate the problem of finding adversarial flows against censoring classifiers 

\noindent as a packet sequence generation task. To solve it, we design Amoeba,\footnote{Our censorship circumvention algorithm's name draws inspiration from the unicellular organism with the same name that is capable of altering its shape. Similarly, our solution alters the shape of network flows.} a novel black-box attack through reinforcement learning,  which learns to craft adversarial flows solely based on the classification results of censoring classifiers, without any further knowledge. This leads to the following \textit{ contributions}:
\begin{enumerate}[leftmargin=*,topsep=2pt]
    \item We make \textbf{no assumption about the underlying model of a censoring classifier}, which {may or may not apply feature engineering} and may not be differentiable (and hence approximating gradients impractical for generating adversarial flows), but instead treat the problem of finding adversarial flows as a process of generating sequences of packets that, when considered together as flows, will be misclassified.
    \item 
    We propose Amoeba, a black-box attack based on \gls{RL} that \textbf{treats the classification results as rewards} and progresses with a policy that maximizes the expected future rewards (return). Our design incorporates a StateEncoder -- a dedicated \gls{NN} that encodes arbitrarily long network flows into fix-sized hidden representations, to help the \gls{RL} agent interpret the context of sequence generation at each timestep.
    \item We evaluate Amoeba on datasets collected using two popular anti-censorship systems, Tor and TLS tunneling; our experimental results indicate that the adversarial flows generated by our Amoeba have \textbf{$\sim$94\% attack success rates, regardless of the type of ML classifier} a censor may deploy. We further show empirically that such adversarial flows are transferable across models with similar architectures. 
    \item {We demonstrate that the Amoeba is \textbf{stable} in different network environments, and \textbf{robust} when receiving noisy and unclear rewards during training.} 
    \item We discuss practical aspects and potential limitations of deploying our Amoeba as a transport layer extension, making the case for its adoption for mainstream censorship circumvention.
\end{enumerate}


\section{Adversarial Models}
\label{sec:threat_model}
As use of ML gains traction in the networking domain, including for \gls{WF} \cite{panchenko2011website,juarez2014critical,panchenko2016website,rimmer2018,sirinam2018deep,sirinam2019triplet} and network intrusion detection \cite{liu2022netsentry,diallo2021adaptive,mirsky2018kitsune}, 
censors are increasingly adopting ML-based classifiers to detect unwanted protocols or traffic associated with banned web services. We consider the most common setting where the censor has full control of the network gateway and enough computational power to examine every network flow traversing it. More precisely, the censor may collect network traffic generated by `forbidden' protocols/web services along with innocuous traffic. A group of statistical features may be extracted from individual flows and fed to a ML classifier for training, as shown in Figure \ref{fig:adv_threat_models}. The censor then deploys the ML model on the gateway to block sensitive flows, e.g., by using and maintaining a blacklist of (src\_ip, src\_port, dst\_ip, dst\_port, protocol) tuples on the firewall. That said, once a traffic flow is recognized as `unwanted' by the censor, the pair of sockets used on the source and destination machine cannot communicate to establish a connection. The censor would not block the destination IP entirely, in order to prevent collateral damage, especially as CDNs increasingly serve thousands of service with the same address \cite{fayed21}. This is a reasonable practical assumption -- for instance, the Great Firewall blocks port numbers instead of IP address when censoring Shadowsocks~\cite{beznazwy2020china}. 

We consider broad scenarios whereby censors need not use deterministic fingerprints in the decision-making process, such as crypto scheme and SNI in TLS handshakes, since these fingerprints can be eliminated easily by fixing the protocol implementation. Also, a censor would not conduct active probing, which is orthogonal to passive observation and outside the scope of our study. 

\begin{wrapfigure}[17]{r}{0.5\columnwidth}
    \centering
    \includegraphics[width=0.5\columnwidth]{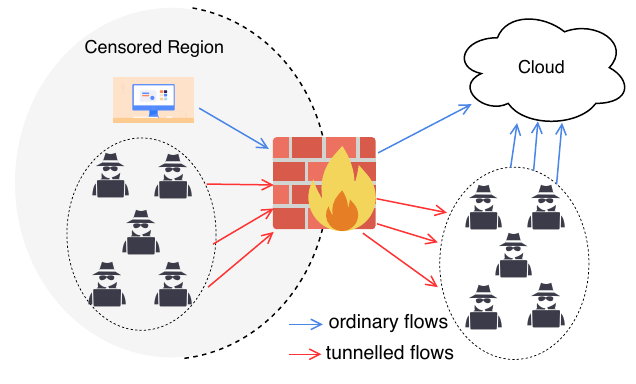}
    \vspace*{-2.5em}
    \caption{Strictest adversarial model considered for subverting Internet censorship. `Attackers' with no knowledge of the censor's tools manipulate packet sizes and inter-packet times based on implicit feedback received (flow permitted or not), to find a tunneled traffic shaping strategy that evades censorship.}
    \label{fig:threat_model}
    \vspace*{-0.5em}
\end{wrapfigure}

We define different capabilities of an `attacker' who attempts to circumvent ML-supported censorship as shown in Figure~\ref{fig:adv_threat_models}. The most rudimentary setting for adversarial attacks is the \textit{white-box model}: the trained censoring classifier is available to the attacker who leverages weight and gradient information to perturb the inputs to the ML model. Under this setting, the attacker also knows the features extracted by the censor, thus perturbations are conducted directly in the feature space instead of on raw flows/packets. A generated adversarial sample is the set of features of a flow, and converting the features back into a legitimate flow is not of this type of attacks' concern. The \gls{CW} attack \cite{carlini2017towards,granados2020realistic} uses projected/clipped gradient descent to find minimal perturbations on the inputs, such that the censoring model would misclassify. \gls{GAN}-based methods \cite{zolbayar2022generating, nasr2021defeating} treat the censoring classifier as the discriminator in a \gls{GAN} and train a generator to produce adversarial samples. 

However, 
given that the censor is unlikely to reveal the feature engineering process, the training technique employed and the architecture of ML models,  we \textit{define a stricter threat model} for adversarial attacks from a realistic perspective, to which we refer as \textit{black-box attack}, as shown in Figure \ref{fig:adv_threat_models}. Assume the attacker has access to a large number of machines with different IP addresses on both sides of the gateway, and can establish connections arbitrarily, as shown in Figure \ref{fig:threat_model}. The adversary may finely manipulate every network flow, by controlling packet sizes and packet inter-arrival times. We assume the manipulated network flows would be examined by the censor and the attacker can reliably infer the the censor's decisions. Under this setting:
\begin{enumerate}[leftmargin=*,topsep=2pt]
    \item The attacker does not know which statistical features the censor may extract from each flow;
    \item The attacker does not know the architecture of the classifying ML model;

    \item The ML model may not be built with \glspl{NN}, but with traditional algorithms, e.g., \gls{SVM} or \gls{DT}, so gradient information is not guaranteed to exist.
\end{enumerate}

This \textit{black-box} setting gives the attacker very limited guidance on how to generate adversarial samples, while the inherent difference between the networking and other domains (e.g., computer vision) precludes the use of existing adversarial input manipulation techniques such as Square Attack~\cite{andriushchenko2020square}, to circumvent censorship. 

\vspace*{-0.25em}
\section{Problem Formulation}
\label{sec:problem_formulation}
Adversarial flows that seek to subvert censorship must accommodate the original payloads and be transmissible in real-world network settings. Thus, we first define a set of practical constraints that adversarial flows must satisfy, then formulate adversarial attacks as a constrained optimization problem, which we solve with a purpose-built \gls{RL} solution.

\textbf{Constraints on Adversarial Attacks:} We represent a network flow by a tuple $S = (P, \Phi)$, where $P$ is a vector of $n$ packet sizes, and $\Phi$ a vector of inter-packet delays, i.e.,
\[
P = [p_{1}^{+}, p_{2}^{-}, ...,p_{n}^{+}],\;\;
\Phi = [\phi_{1}, \phi_{2}, ...,\phi_{n}].
\]
Superscript `$+$' indicates packets transmitted from client to server, and `$-$' vice versa. An adversarial 
sample can alter each packet size by padding or truncation, and can delay packets to deceive ML classifiers. However, the attacker must ensure that bidirectional payloads in the original flows are transmitted in the correct order. Denote $\tilde{S} = (\tilde{P}, \tilde{\Phi})$ as the adversarial version of flow $S$, where
\begin{align*}
    \tilde{P} = [\tilde{p}_{1, 1}^{+}, ..., \tilde{p}_{1, k_{1}}^{+}, \tilde{p}_{2, 1}^{-}, ..., \tilde{p}_{2, k_{2}}^{-}, ..., \tilde{p}_{n, 1}^{+}, ..., \tilde{p}_{n, k_{n}}^{+}], 
    \tilde{\Phi} = [\tilde{\phi}_{1, 1}, ..., \tilde{\phi}_{1, k_{1}}, \tilde{\phi}_{2, 1}, ..., \tilde{\phi}_{2, k_{2}}, ..., \tilde{\phi}_{n, 1}, ..., \tilde{\phi}_{n, k_{n}}].
\end{align*}
The sub-sequence $[\tilde{p}_{i, 1}, ..., \tilde{p}_{i, k_{i}}]$ represents the adversarial manipulation of original packet sizes $p_{i}$, with $\{k_{1}, ..., k_{n}\}$ denoting the lengths of all sub-sequences. Since we allow for both packet truncation and padding, the length of an adversarial flow can be larger than that of the original, i.e., $|\tilde{P}| \geq |P|$, though the following constraint on packet sizes must be satisfied:
\begin{equation}
\label{eq:packet_constraint}
    \sum_{j=1}^{k_{i}} \tilde{p}_{i, j} \geq p_{i}, \; \forall i \in [1, n],
\end{equation}
which ensures that each original packet can be transmitted without data loss. It is straightforward to derive constraints on timestamps: 
\begin{equation}
\label{eq:time_constraint}
    \tilde{\phi}_{i, 1} \geq \phi_{i}, \tilde{\phi}_{i, j} \geq 0, \; \forall j \in [1, k_{i}], i \in [1, n].
\end{equation}

\textbf{Finding Adversarial Samples:} Let $e(\cdot)$ be a feature extraction function that takes an arbitrary flow $S$ and outputs $d$-dimensional features. Denote $f: \mathcal{R}^{d} \rightarrow [0, 1]$ a binary classifier. $f$ can be a neural network using a sigmoid as the  activation function in the final layer, so its output $y = f (e(S))$ is a real number between 0 and 1. Alternatively, $f$ can be a traditional ML algorithm (SVM, decision tree, etc.), which directly outputs discrete classification results ($\{0, 1\}$). If using a NN-based classifier, the censor would use a decision~function
\[
C(y) = 
\begin{cases}
    1, & \text{if } y\geq 0.5;\\
    0,              & \text{otherwise}.
\end{cases}
\]
That said, if the predicted score is smaller than 0.5, the flow is to be blocked. A flow $\tilde{S}$ is regarded as an adversarial version of $S$ if $C(f(e(\tilde{S}))) = 1$. 
The task of finding $\tilde{S}$ can be rephrased as a constrained optimization problem:
\begin{align*}
 \max \; C(f(e(\tilde{S}))) 
\text{  s.t.}  \sum_{j=1}^{k_{i}} \tilde{p}_{i, j} \geq p_{i}, \;\; \tilde{\phi}_{i, 1} \geq \phi_{i}, \;\; \tilde{\phi}_{i, j} \geq 0, \forall j \in [1, k_{i}], \;\; \forall i \in [1, n].    
\end{align*}


\textbf{Are upper bound constraints necessary?} Different from the computer vision domain, we do not impose upper bound constraints on both payload and timing. The adversarial examples $\tilde{x}$ of \textit{an image} $x$ must satisfy an $l_{p}$-norm bound, i.e., $||\tilde{x} - x||_{p} \leq \epsilon$ \cite{andriushchenko2020square}, because $\tilde{x}$ should not tamper with the semantics of the original image $x$ from a human perspective. An image of a panda should still `look like' a panda after adversarial perturbation. However, in the networking domain, as long as the original payload is transmitted, and the sender and the recipient can interpret messages identically, the semantics remain the same. Therefore, minimizing data overhead and timing delays are not hard constraints for the problem we solve, but optional requirements that users may have in order to prevent performance degradation, for which we also offer a solution in Section \ref{sec:env_setup}. 

\begin{figure*}[t]
    \centering
    \hspace*{-0.2cm}
    \includegraphics[width=0.9\linewidth]{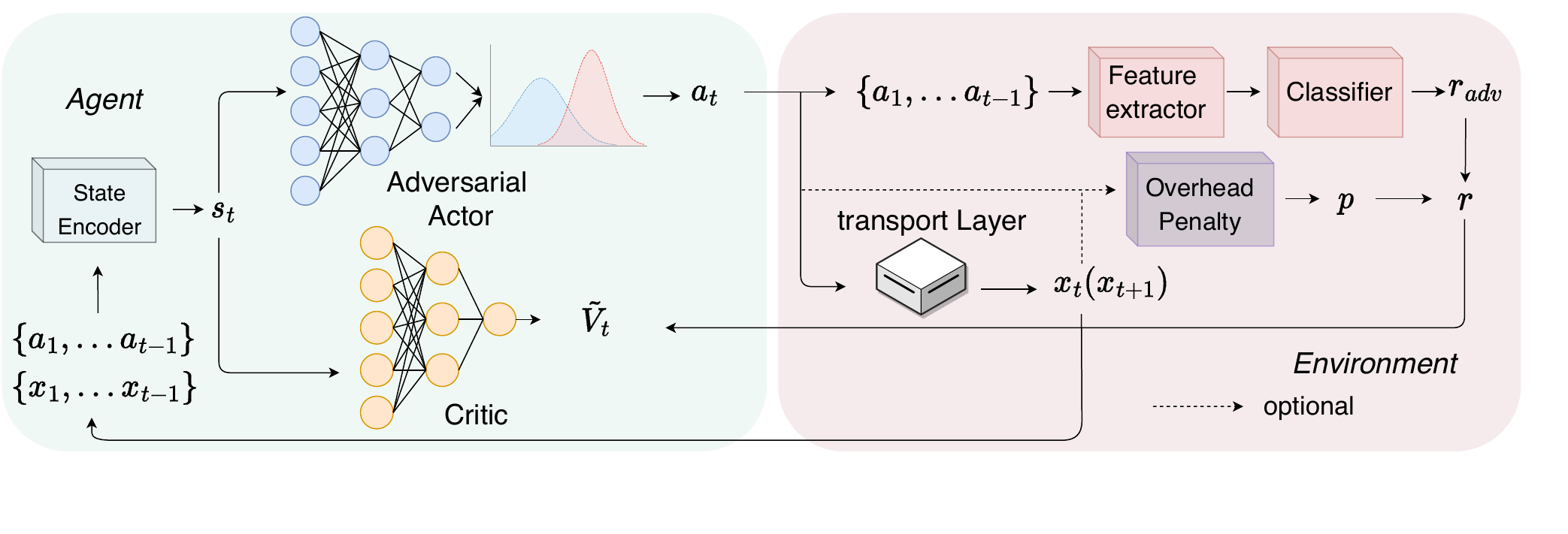}
    \vspace*{-3em}
    \caption{The architecture of the proposed adversarial reinforcement learning algorithm -- Amoeba. }
    \label{fig:addpg}
    \vspace*{-1.5em}
\end{figure*}

\vspace*{-0.5em}
\section{Amoeba Design}
\label{sec:addpg}
Traditional adversarial attacks do not comply with the specifics of network flows, because (1) the length of adversarial samples are variable, according to each flow, and the optimal length is unknown; and (2) one should be able to send adversarial samples packet-by-packet, 
whereas existing attacks 
generate the feature set of an entire flow at once, without considering the practicalities of transmission. 

Instead, \textbf{we regard finding adversarial versions of network flows as a sequence generation process}, which takes an input (a packet and the associated timestamp in the original flow) at each timestep and outputs adversarial manipulations of that input. The adversarial packets can be transmitted in almost real-time, rather than waiting for the entire flow to finish first. \textcolor{black}{Each packet in a flow should be morphed to maximize the chances that the complete flow in the future will be misclassified, which requires an algorithm to look ahead of time and progresses through binary signals received from the censor.}
Given these requirements, \gls{RL} is particularly well-suited to our task, where we treat the output of~the censoring classifier as reward that~guides a \gls{RL} agent to learn a packet sequence generation policy. 
We design Amoeba to generate adversarial flows that circumvent censorship. Amoeba models censor decisions as a reward function, and trains an agent to interact with the censor in discrete timesteps. At each step $t$, the agent receives a packet from the transport layer, takes an adversarial action (effectively modifying the size and timing of the packet), and obtains a reward based on how good that action is. The agent aims to maximize the future rewards when generating adversarial samples. Note that \textbf{Amoeba does not change the implementation of any existing protocol} in terms of handshake, error handling and acknowledgment, but simply alters the `shape' of each packet with payload to deceive \gls{ML} classifiers. In other words, an adversarial TCP flow is still a legitimate TCP flow.
Amoeba comprises four major elements: Network Environment, State Encoder, an Adversarial Actor and a Critic (see Figure~\ref{fig:addpg}). Next, we provide a \gls{RL} primer, before diving into our solution.

\subsection{RL Primer}
\label{sec:rl_summary}
Our algorithm takes a reinforcement learning approach with an agent interacting with the environment in discrete timesteps. 
At step $t$, the environment gives an \textit{observation} $x_{t} = (p_{t}, \phi_{t})$, representing an original packet to send with size $p_{t}$ and inter-packet delay $\phi_{t}$. For each flow, the actor maintains a vector of previous observations $[x_{1}, x_{2},..., x_{t-1}]$, as well as a vector of previous actions $[a_{1}, a_{2}, ..., a_{t-1}]$, with each \textit{action} $a_{i} = (\tilde{p}_{i}, \tilde{\phi}_{i})$ representing the manipulation of an original packet $x_{i}$. In this paper, we use \textit{actions} and \textit{adversarial packets} interchangeably. The \textit{state} at step $t$ is the history of both the observations and the actions, i.e., $s_{t} = (x_{1}, a_{1}, ..., x_{t-1}, a_{t-1}, x_{t})$. Note that $s_{t} \neq x_{t}$, because the actor needs a broad understanding of the current environment based on what has been generated up to that point. The actor parameterized by $\theta$ maps a state to a probability distribution over the actions $\pi_{\theta}(s_{t})$. An action is randomly sampled $a_{t} \sim \pi_{\theta}(s_{t})$, and given to the environment, leading to a \textit{reward} $r(s_{t}, a_{t})$ and the next observation $x_{t+1}$. An \textit{episode} $\tau$ indicates the entire process of generating an adversarial sample given a flow, i.e. $\tau := (s_{1}, a_{1}, ..., s_{T}, a_{T})$. The aim of the actor is to select actions at every timestep in a way that maximizes the total future rewards:
\[
\max_{\theta} \mathbb{E}_{\tau \sim p_{\theta}(\tau)}[\sum_{t=1}^{T} r(s_{t}, a_{t})].
\]
The above problem can be solved by iteratively updating $\theta$ \cite{sutton1999policy}: 
\begin{align*}
    \theta_{k+1}  = \theta_{k} + \alpha \mathbb{E}_{t} \left[ \nabla_{\theta} \log \pi_{\theta}(a_{t} | s_{t}) Q^{\pi}(s_{t}, a_{t}) \right],
\end{align*}
where $\alpha$ represents step size, and $Q^{\pi}(s_{t}, a_{t})$ is known as the \textit{action-value function} that produces the discounted total future reward. Approximating $Q$ values directly suffers from high variance in practice. Thus, a baseline is always subtracted from $Q$ while keeping the objective unbiased \cite{sutton1999policy}:
\begin{align}
\label{eq:actor_loss}
    \theta_{k+1}  = \theta_{k} + \alpha \mathbb{E}_{t} \left[ \nabla_{\theta} \log \pi_{\theta}(a_{t} | s_{t}) A^{\pi}(s_{t}, a_{t}) \right]
\end{align}
in which \textit{Advantage} $A(s_{t}, a_{t}) = Q^{\pi}(s_{t}, a_{t}) - V^{\pi}(s_{t})$. Here, the second term is called the \textit{state-value function} $V^{\pi}(s_{t}) = \mathbb{E}_{a_{t} \sim \pi}[Q(s_{t}, a_{t})]$, which represents the expected future reward from step $t$, and $A(s_{t}, a_{t})$ intuitively indicates how much better the current action $a_{t}$ is than the average.

\subsection{Environment}
\label{sec:env_setup}
The network environment offers observations and rewards, given new actions. 

\paragraph{Generating Observations} In practice, observations (packets) originate from the buffer in the transport layer. When there is no traffic obfuscation in place, the payload in the buffer would be encapsulated in packets and transmitted immediately. However, to generate adversarial samples, the payload cannot be sent directly but should be passed through the adversarial actor, which decides appropriate packet sizes and timings, such that the $Q$ value can be maximized. 

Therefore, as the first step we use a transport layer emulator that reads a payload with $p_{i}$ bytes from the buffer as the vanilla transport layer does. To adversarially manipulate this packet (observation),  $x_{t} = (p_{i}, \phi_{i})$ is given to the agent, which morphs the packet based on a given policy $\pi$, truncating or adding padding to it along with some delays. Both truncation and padding are supported to expand the action space that the agent can explore, and thereby create adversarial flows with more variability. Only supporting either operation may result in the failure of generating adversarial flows. For example, an attack by only padding cannot circumvent censoring models \cite{rimmer2018,sirinam2018deep} that leverage directional features, since padding only changes the size of each packet but the packet directions in a flow remain the same after morphing; attacks by only truncating may hardly protect protocols with fixed payload unit size such as Tor cells, given that censoring can easily recover by summing the packet sizes in the same direction.
Since we allow for truncation, it is possible that the adversarial packet (action) $\tilde{a}_{t} = (\tilde{p}_{i}, \tilde{\phi}_{i})$ is smaller than the original one, leaving $p_{i} - \tilde{p}_{i}$ byte payload to send. In that case, the emulator does not read more payload from the buffer, but generates a second adversarial packet by giving the agent $x_{t+1} = (p_{i} - \tilde{p_{i}}, \phi_{i+1})$. Such operation is repeated until the remaining payload is fully sent, and then the emulator reads more payload from the buffer. For example, assume the agent truncates an original packet $n$ times, the list of the observations sent to the agent and the list of actions would be:
\begin{align*}
   [(p_{i}, \phi_{i}), (p_{i}-\tilde{p}_{i, 1}, \phi_{i+1}), ..., (p_{i} - \sum_{j=1}^{n-1}\tilde{p}_{i, j}, \phi_{i+n-1})], \text{ and  }
    [(\tilde{p}_{i, 1}, \tilde{\phi}_{i, 1}), (\tilde{p}_{i, 2}, \tilde{\phi}_{i, 2}), ..., (\tilde{p}_{i, n}, \tilde{\phi}_{i, n})]. 
\end{align*}
Padding occurs if the final adversarial packet is larger than the input size, i.e., $\tilde{p}_{i, n} > p_{i} - \sum_{j=1}^{n-1}\tilde{p}_{i, j}$.
Observe that \textbf{the emulator satisfies the constraint on packet sizes (Eq. \ref{eq:packet_constraint}) by design}, so that the adversarial actor does not have to consider it while learning the policy. Also, the observation $x_{t}$ and the associated packet size $p_{i}$ do not share the same subscript because the emulator may read from buffer once, but uses multiple timesteps to send the payload, due to truncation.

\paragraph{Reward Function Design} The reward function evaluates how good an action $a_{t}$ is under the current state $s_{t}$. Since our aim is to find adversarial network flows, the reward should first reflect the judgment of the censor, i.e., $C(f(e(\cdot)))$. There are two standard strategies to assign rewards for each action-state pair. The first is not assigning intermediate rewards while the sequence is being generated, but only assigning a final reward when the episode terminates. One typical example is AlphaGo \cite{silver2017mastering}, which assigns either $+1$ or $-1$ when a round of go game ends. The other strategy is to give a reward at each timestep, which was adopted for cartpole or Mario game play. The first strategy might seem suitable for our task, since all the intermediate actions should serve the final aim, that is, the adversarial flow as a whole should be misclassified. However, this would imply that the environment knows in advance when a flow will terminate, so it defers a reward until the last packet. In reality, a flow may terminate at an arbitrary timestep due to different communication purposes or network status. Note that in our adversarial model, attackers can control each packet, meaning that they can also terminate a flow at any point. {In other words, we consider it possible for the censor to make a classification decision at any timestep, as if this is the last in an episode.}

Formally, consider $a_{t} = (\tilde{p}_{i, n}, \tilde{\phi}_{i, n})$ at timestep $t$ is generated by the attacker given $x_{t} = (p_{i} - \sum_{j=1}^{n-1}\tilde{p}_{i, j}, \phi_{i+n-1})$ and sent over the network. The censor already witnesses $\mathbf{a}_{1:t} = [a_{1}, a_{2}, ..., a_{t}]$. Thus, the reward regarding distinguishability is defined as:
\[
r(s_{t}, a_{t})_{adv} =  - C(f(e(\mathbf{a}_{1:t}))).
\]

Besides, we also consider extra penalties in terms of data overhead and time delays. One may expect the adversarial sample is as close to the original flow as possible, i.e., introducing the smallest padding and delays, which would do minimal harm to the application performance. We therefore introduce a data overhead penalty and a time overhead penalty: 
\begin{align*}
    p(s_{t}, a_{t})_{data} &= \begin{cases}
    p_{i} - \sum_{j=1}^{n} \tilde{p}_{i, j} + \lambda_{split}n, & \text{if $\tilde{p}_{i, n} < p_{i} - \sum_{j=1}^{n-1} \tilde{p}_{i, j}$};\\
    \sum_{j=1}^{n} \tilde{p}_{i, j} - p_{i}, & \text{otherwise}. 
\end{cases}
\end{align*}

When the size of the adversarial packet at timestep $t$ is smaller than that of the original packet, the penalty is proportional to the number of truncations $n$ plus the remaining bytes to send. When padding occurs ($\tilde{p}_{i, n} > p_{i} - \sum_{j=1}^{n-1}\tilde{p}_{i, j}$), the penalty is linear in the extra bytes to send. We do not use symmetric penalties for the two circumstances, because we find empirically that Amoeba is inclined to truncate packets into multiple instances of minimal size. Thus, we discourage this behavior by assigning an extra penalty when packet truncation occurs. The penalty for time delays is straightforward: $ 
p(s_{t}, a_{t})_{time} = \tilde{\phi}_{i, n} - \phi_{i+n-1}.$. The expression of the reward function thus becomes:
\[
r(s_{t}, a_{t}) = r(s_{t}, a_{t})_{adv} - \lambda_{d} p(s_{t}, a_{t})_{data} - \lambda_{t} p(s_{t}, a_{t})_{time},
\]
where $\lambda_{split}$, $\lambda_{d}$ and $\lambda_{t}$ are hyperparameters that balance each component.

\subsection{Adversarial Actor \& Critic}
As mentioned in Section \ref{sec:rl_summary}, the state at timestep $t$ is the history of the observations and the actions, meaning that the length of the state would vary as $t$ increases. However, if the agent is built with non-recurrent neural networks, such as \gls{MLP}, it requires inputs of fixed size. To overcome this problem, we design StateEncoder, a two-layer, pre-trained \gls{GRU} that encodes an arbitrary long network flow to a fixed-size hidden representation. The pretraining and the performance of the StateEncoder are documented Appendix \ref{sec:stateencoder} and \ref{sec:perf_stateencoder}.

The adversarial actor aims to pick an optimal action at each timestep, such that the future rewards can be maximized. However, the action space for packet sizes is overwhelmingly large, i.e., 1,448 discrete actions for TCP and 16,384 for TLS, while the action space for time delays is infinite. Thus, we first treat both packet sizes $p$ and time delays $\phi$ as continuous, and discretize them when the actor makes a choice. For example, for the TCP layer, we let the actor choose an action $(p_{i}, \phi_{i}),  p_{i} \in [-1, 1], \phi_{i} \in [0, 1]$, and then discretize the packet size by $int(p_{i} \times 1,460)\;byte$ and the time delay $int(\phi_{i} * max\_delay)\;ms$, where $max\_delay$ indicates the maximum allowed delay for a packet. Note that packet sizes can be negative to represent backward traffic.

We follow an actor-critic design where the actor network $\pi_{\theta}(\cdot)$ approximates the best action given a state, and a critic network $V_{c}(\cdot)$ estimates the state value. The two networks are parameterized by $\theta$ and $c$ respectively. Specifically, the learning objective of the actor is as described by Eq. (\ref{eq:actor_loss}). The critic network aims to approximate the state-value by minimizing the Mean Squared Error between estimated values and the discounted future rewards ($R_{t}$):
\begin{align}
    \begin{split}
    \min_{c} \mathbb{E}_{t}[(V_{c}(s_{t}) - E_{a \sim \pi}[Q(a_{t}, s_{t})])^{2}]
     \approx \mathbb{E}_{t}[(V_{c}(s_{t}) - R_{t})^{2}].
    \end{split}
\end{align}
\vspace{-0.25em}
In practice, we set $\pi_{\theta}$ and $V_{c}$ as \glspl{MLP} and find this network structure to be effective in our task. The adversarial actor has two output units: packet size $\tilde{p}$ and inter-packet delay $\tilde{\phi}$. \textbf{To satisfy the time constraint on inter-packet delays in Eq.~\ref{eq:time_constraint}}, we let $\pi_{\theta}$ output a value $\Delta_{\phi}$ representing how much extra delay should be added to each packet apart from the existing delay $\phi$ provided by the environment, i.e., $\tilde{\phi} = \phi + \Delta_{\phi}$.

\subsection{Optimization}
Optimizing \gls{RL} algorithms is challenging due to high variance among trajectories and the trade-offs between exploration and exploitation that need to be achieved. A few techniques are widely used to stabilize the training process, speed up convergence, and ensure the networks are differentiable, which we also adopt in training our agent, including (1) surrogate objective function \cite{schulman2017proximal}; (2) reparameterization trick, and (3) parallel rollout \cite{schulman2017proximal}. Interested readers can refer to Appendix \ref{sec:optimization_details}. The full training algorithm is detailed in Algorithm \ref{alg:amoeba_trianing}.

\begin{wrapfigure}[35]{L}{0.6\textwidth}
\vspace{-1.5em}
\begin{minipage}{0.6\textwidth}
\setlength{\textfloatsep}{5pt}%
\begin{algorithm}[H]
\small
  \caption{The training algorithm of Amoeba}
  \label{alg:amoeba_trianing}
  \begin{algorithmic}[1]
    \Inputs{$\lambda_{split}: =$ packet truncation overhead coefficient \\$\lambda_{d}: =$ data overhead coefficient; $\gamma :=$ discount factor \\ 
    $\lambda_{t}: =$ time delay coefficient; $N:=$ number of environments\\
    $T:=$ the length of each rollout in the environment; }
    \Initialize{
    Initialize $\pi_{\theta}$ and $V_{c}$ via Xavier initialization \cite{glorot2010understanding}\\
    Obtain StateEncoder $\mathcal{E}$ from Algorithm \ref{alg:stateencoder_trianing}\\
    Initialize N $Env$, each of which is provided a feature extractor $e(\cdot)$ and a pretrained classifier $f(\cdot)$ \\
    Initialize a rollout buffer with size $N \times T$
    }
    \While{not converged}
        \State {Sample $N \times T$ observations by interacting $\pi_{\theta}$ with $N$ $Env$}
        \For{Each observation $x_{t}$, action $a_{t}$ and reward $r_{t}$}
            \State {Let $\mathbf{x}_{1:t} := \{ x_{1}, ... x_{t}\}, \mathbf{a}_{1:t} := \{ a_{1}, ... a_{t}\}$}
            \State {Generate the state representation \\ by $s_{t} = \mathcal{E}(\mathbf{x}_{1:t}) || \mathcal{E}(\mathbf{a}_{1:t})$}
            \State {Compute $\hat{A}_{t}=\sum_{l=0}^{\infty} (\gamma \lambda)^{l} [r_{t+l} + \gamma V(s_{t+l+1}) - V(s_{t+l})]$}
            \State {Compute Return $R_{t} = \hat{A}_{t} + V_{c}(s_{t})$}
        \EndFor
        \State {Store each $(s_{t}, a_{t}, r_{t}, \hat{A}_{t}, R_{t})$ in the rollout buffer and split them into $K$ mini-batches $\{ \mathcal{D}_{1}, .., \mathcal{D}_{K}$ \}.}
        \State {Set $\pi_{\theta_{old}} \gets \pi_{\theta}$}
        \For{$k=1,K$}
                \State {Compute $I_{t}(\theta) = \frac{\pi_\theta\left(a_t \mid s_t\right)}{\pi_{\theta_{old}}(a_{t} | s_{t})}$}
                \State {Update $\theta$ via \\
                $ \nabla_{\theta}  \frac{1}{|\mathcal{D}_{k}|}\sum \left[\min (I_{t}(\theta) \hat{A}_{t}, clip(I_{t}(\theta), 1-\epsilon, 1+\epsilon) \hat{A}_{t}) + H_{\pi_{\theta}}(a_{t}) \right]$}
                \State {Update $c$ via $ -\nabla_{c} \frac{1}{|\mathcal{D}_{k}|} \sum \left( V_{c}(s_{t}) - R_{t}\right)^2 $}
        \EndFor
    \EndWhile
    \State {\Return {$\mathcal{E}$}}
    \end{algorithmic}
\end{algorithm}
\end{minipage}
\vfill
\end{wrapfigure}

\section{Experiments}
\label{sec:exp}
In this section, we empirically evaluate the effectiveness of applying Amoeba on two popular types of anti-censorship systems, namely Tor network and generic TLS tunneling:
\vspace{-0.25em}
\begin{enumerate}[leftmargin=*]
    \item Tor Network is a anonymity system that utilizes relay nodes with onion protocol to conceal user location and prevent network surveillance \cite{tor}. The traffic routed inside the Tor network is encrypted by TLS and only the exit node has access to the original traffic, which is forwarded to the real destination. However, Tor is proven to be distinguishable by ML classifiers due to the fixed-size cells of the onion protocol \cite{wang2015seeing}.
    \item V2Ray is a generic TLS tunneling tool that tunnels arbitrary TCP/UDP packets inside TLS connections \cite{v2ray}. Users of this type of systems usually do not demand anonymity but only seek to bypass firewalls. Thus, these tools are widely used in countries that employ censorship, such as China. We use V2Ray as the supporting tunneling system rather than its alternatives \cite{trojan, lantern, naiveproxy}, given that it has the largest community of both maintainers and users, and it is also widely supported by 3\textsuperscript{rd} party clients across multiple platforms \cite{v2ray-client}.
\end{enumerate}
\vspace{-0.25em}

Both system types are vulnerable to ML classifiers due to the fact that the statistical features of the tunneled flows deviate from real TLS/HTTPS traffic. 
\vspace{-0.25em}

\subsection{Censoring Classifiers}
We adopt a range of state-of-the-art traffic analysis models as censoring classifiers: 

\textbf{\gls{DF} \cite{sirinam2018deep}} is a state-of-the-art \gls{CNN}-based deep learning model that automatically extracts features from raw network flows and performs~\glsfirst{WF}. 

\textbf{\gls{SDAE} \cite{rimmer2018}} follows a \gls{MLP}-based encoder-decoder architecture to 
extract latent features from network flows directly for \gls{WF}. 

\textbf{\gls{LSTM} \cite{rimmer2018}} is a multi-layer recurrent neural network that takes arbitrary long network flows as input to perform \gls{WF}. \gls{LSTM} is designed to learn long-term dependencies, and therefore can better interpret timeseries data such as consecutive packets.  

\textbf{CUMUL \cite{panchenko2016website}} separates different classes of data by using \gls{SVM} with a radial basis function (RBF) kernel to find the hyperplane that maximizes the margin between classes.

The original versions of \gls{DF}, \gls{SDAE} and \gls{LSTM} are fed with packet directions only (i.e., $(-1, 1)$), and vanilla CUMUL leverages the cumulative representation of network traces without timing features. For consistence, we tailor these classifiers to utilize the flow representation in Sec.~\ref{sec:problem_formulation} as input. That said, these classifiers do not need an external feature extractor. 

\textbf{Tree-based models \cite{barradas2018effective}:} Traditional \gls{ML} models, such as \gls{DT} and \gls{RF}, exhibit promising performance in detecting multi-media tunneling protocols . Tree-based approaches possess better interpretability compared to DL models, since the decision-making process can be visualized as a set of tree-like rules. We follow \cite{barradas2018effective} to extract 166 features from each network flow, covering bi-directional packet/timing statistics, burst behaviors, percentile features and flow-level information, and use them to train the \gls{DT}/\gls{RF}.

\vspace*{-0.5em}
\subsection{Adversarial Attack Benchmarks}
We choose three advanced \textit{white-box} adversarial attacks as benchmark algorithms for our evaluation:

\textbf{\glsfirst{CW} Attack \cite{carlini2017towards}} uses projected gradient descent to find minimal perturbations on the inputs, while maximizing the probability of the inputs being misclassified. The \gls{CW} attack iteratively queries the classifier for a single input, until an adversarial sample is found.

\textbf{NIDSGAN \cite{zolbayar2022generating}} regards the censoring classifier as the discriminator in a \gls{GAN} architecture, and trains a generator to learn minimal perturbation patterns needed to fool the discriminator. The generator directly produces adversarial samples given inputs, without needing  iterative updates.

\textbf{\gls{BAP} \cite{nasr2021defeating}} also aims at training generator-like \glspl{NN}, but is more flexible in allowing inserting packets into a given flow, i.e., the length of an adversarial sample is not always identical to the input, posing larger difficulties for censoring classifiers.   

We do not consider black-box benchmark algorithms \cite{andriushchenko2020square, brendel2017decision, chen2020hopskipjumpattack}, since existing ones are infeasible under our threat model where feature extraction is performed (see Figure \ref{fig:adv_threat_models}). We implement the \gls{NN}-based classifiers, \gls{CW} attack, NIDSGAN, \gls{BAP} and Amoeba in Pytorch \cite{paszke2019pytorch}, and import the rest of the classifiers from the scikit-learn package in Python. Detailed hyperparameter selection is documented in Appendix \ref{sec:hyper_selectioin}.

\vspace*{-0.5em}
\subsection{Evaluation Metrics}
To evaluate the effectiveness of our solution against ML classifiers, we measure their accuracy and F1 score metrics, which are based on True Positives (TP), False Positives (FP), True Negatives (TN) and False Negatives (FN): $accuracy = (TP + TN)/(TP + TN + FP + FN)$, and $F1 = 2 \times (precision \times recall)/(precision + recall)$, where $precision$ $=$ $TP/(TP+FP)$ and $recall$ $=$ $TP/(TP+FN)$. Accuracy indicates the proportion of samples correctly classified, and F1 score computes the harmonic mean between precision and recall. The former represents how likely an algorithm would give true alarms, and the latter indicates how sensitive an algorithm is towards positive samples. 

We use also use another three metrics to evaluate Amoeba, namely Attack Success Rate (ASR), i.e., the percentage of adversarial samples being misclassified, $data\;overhead = padding / (original\;payload$ $+ padding)$ and $time\;overhead = delays/(delays + total\; transmission$ $\; time)$,
in which total transmission time is the time difference between the last and first packet in a flow.

\subsection{Data Collection \& Training Procedure}
\label{sec:data_split}
We collect two real-world datasets to evaluate our approach. Specifically, we set up a Tor client on a campus machine running Ubuntu 22.04, and a Tor bridge on a Google Cloud E2 instance running Ubuntu 22.04. TCP segmentation offload is disabled on both machines. The same setup is employed for a V2Ray client and V2Ray proxy server. We consider the censor sits between the Tor (or V2Ray) client and the first relay node (or V2Ray server) and distinguishes sensitive flows. To collect a realistic Tor dataset for evaluation, we crawl the landing pages of Alexa top 25,000 websites with and without Tor network respectively
(\texttt{Tor Dataset}). We use tshark to group packets into TCP flows and extract packet sizes and associated timestamps, where backward packet (server-to-client) sizes are represented with negative numbers to preserve the transmission direction.
The same operation is repeated with and without the V2Ray tunnel, named \texttt{V2Ray Dataset}. Different from the \texttt{Tor Dataset}, we utilize tshark to group packets into TLS flows, and extract TLS record sizes and timestamps. For this dataset, we consider the censor conducts deep packet inspection up to the TLS layer and extracts features from TLS flows instead of TCP flows. The maximal TLS record is 16 KB, i.e., Amoeba is required to explore a much larger action space.

Each dataset is separated into a \textit{clf\_train\_set} ($40\%$), an \textit{attack\_train} \textit{\_set} ($40\%$), a \textit{validation\_set} ($10\%$) and a \textit{test\_set} ($10\%$). We use the \textit{clf\_train\_set} to train censoring classifiers, which are then evaluated on the \textit{test\_set}. After that, each trained censoring classifier is deployed in the Environment in Figure \ref{fig:addpg} to generate rewards. We use the \textit{attack\_train\_set} to train Amoeba instead of using \textit{clf\_train\_set}, because the attacker may have no access to the dataset owned by the censor in practice. The \textit{validation\_set} is utilized to tune the hyperparameters of Amoeba. After training, Amoeba and the benchmark algorithms are evaluated on the \textit{test\_set} against the trained censoring classifiers. To facilitate the reproducibility of our results, we make available our data collection configurations, datasets and source code at \url{https://github.com/Mobile-Intelligence-Lab/Amoeba}.

\vspace*{-1em}
\subsection{Evaluation}

\begin{table*}[t]
\vspace{-1em}
\centering
\small
\bgroup
\setlength{\tabcolsep}{2pt}
\begin{tabular}{c|c|cc|ccc|ccc|ccc|ccc}
\Xhline{2\arrayrulewidth}
\multirow{3}{*}{Dataset} & Attack         & \multicolumn{2}{c|}{None} & \multicolumn{3}{c|}{C\&W}    & \multicolumn{3}{c|}{NIDSGAN}   & \multicolumn{3}{c|}{BAP}   & \multicolumn{3}{c}{Amoeba}         \\
                         & Threat Model   & \multicolumn{2}{c|}{\textit{None}} & \multicolumn{3}{c|}{\textit{white-box}}      & \multicolumn{3}{c|}{\textit{white-box}}  & \multicolumn{3}{c|}{\textit{white-box}}    & \multicolumn{3}{c}{\textit{black-box}} \\
                         & \begin{tabular}{@{}c@{}}Censoring \\ Alg.\end{tabular} & F1       & \begin{tabular}{@{}c@{}}Accu- \\ racy\end{tabular}      & \begin{tabular}{@{}c@{}}ASR \\ (\%)\end{tabular}           & \begin{tabular}{@{}c@{}}DO \\ (\%)\end{tabular}          & \begin{tabular}{@{}c@{}}TO \\ (\%)\end{tabular}         & \begin{tabular}{@{}c@{}}ASR \\ (\%)\end{tabular}           & \begin{tabular}{@{}c@{}}DO \\ (\%)\end{tabular}          & \begin{tabular}{@{}c@{}}TO \\ (\%)\end{tabular}     & \begin{tabular}{@{}c@{}}ASR \\ (\%)\end{tabular}           & \begin{tabular}{@{}c@{}}DO \\ (\%)\end{tabular}          & \begin{tabular}{@{}c@{}}TO \\ (\%)\end{tabular}    & \begin{tabular}{@{}c@{}}ASR \\ (\%)\end{tabular}           & \begin{tabular}{@{}c@{}}DO \\ (\%)\end{tabular}          & \begin{tabular}{@{}c@{}}TO \\ (\%)\end{tabular}       \\ \hline
\multirow{7}{*}{\texttt{Tor}}     & SDAE            & 0.99        & 0.99             & 88.34             & 21.60            & 0.00          & 30.75           & 20.00            & 4.25        & 84.72            & 22.95           & 21.21    & 89.0            & 64.8          & 8.72         \\
                         & DF            & 0.99        & 0.99            & 97.88             & 26.68            & 23.94          & 94.13             & 31.8           & 7.28    & 89.46            & 35.95           & 12.49      & 87.5            & 59.0           & 7.79         \\
                         & LSTM        & 0.99        & 0.99             & 90.49             & 86.64            & 8.37          & 97.88             & 19.09            & 3.54     & 93.86            & 38.88           & 18.65     & 98.2            & 58.1          & 6.26         \\
                         & DT             & 1.00        & 1.00             & \multicolumn{3}{c|}{\multirow{3}{*}{N/A}} & \multicolumn{3}{c|}{\multirow{3}{*}{N/A}}  & \multicolumn{3}{c|}{\multirow{3}{*}{N/A}}  & 96.5            & 39.0           & 5.69         \\
                         & RF             & 1.00        & 1.00             & \multicolumn{3}{c|}{}                     & \multicolumn{3}{c|}{}         & \multicolumn{3}{c|}{}           & 92.0            & 39.1           & 3.73         \\
                         & CUMUL            & 0.99        & 0.99             & \multicolumn{3}{c|}{}                     & \multicolumn{3}{c|}{}            & \multicolumn{3}{c|}{}         & 93.0            & 44.5           & 6.55         \\ \hline
\multirow{7}{*}{\texttt{V2ray}}   & SDAE            & 0.99        & 0.99             & 99.54             & 25.24            & 24.88          & 26.04            & 22.99            & 20.23    & 79.92            & 26.76           & 5.99      & 93.8           & 43.2           & 5.49         \\
                         & DF            & 0.99        & 0.99             & 84.33             & 49.31            & 49.89          & 95.44             & 22.9          & 9.17     & 62.57            & 25.13           & 0.00     & 96.8            & 46.1           & 7.45         \\
                         & LSTM        & 0.99        & 0.99             & 96.61             & 16.10            & 2.51          & 93.32             & 38.44             & 15.23             & 91.56            & 16.98           & 29.78  & 89.2    & 7.73           & 1.46         \\
                         & DT             & 1.00        & 1.00            & \multicolumn{3}{c|}{\multirow{3}{*}{N/A}} & \multicolumn{3}{c|}{\multirow{3}{*}{N/A}} & \multicolumn{3}{c|}{\multirow{3}{*}{N/A}}   & 97.2            & 40.2           & 8.44         \\
                         & RF             & 1.00        & 1.00             & \multicolumn{3}{c|}{}                     & \multicolumn{3}{c|}{}        & \multicolumn{3}{c|}{}             & 99.4           & 53.97           & 8.30         \\
                         & CUMUL            & 0.99        & 0.99             & \multicolumn{3}{c|}{}                     & \multicolumn{3}{c|}{}            & \multicolumn{3}{c|}{}         & 96.4           & 51.6           & 10.48         \\ \Xhline{2\arrayrulewidth}
\end{tabular}
\egroup
\caption{Performance of different classifiers in detecting Tor flows without attack; performance of Amoeba in crafting adversarial flows. For comparison, we also show the \gls{ASR} of \gls{CW}, NIDSGAN, and \gls{BAP} attacks under different threat models (DO -- data overhead; TO -- time overhead). The estimated values reported represent the maximal perturbation allowed for data and timing features respectively.}
\label{table:tor_results}
\vspace{-3em}
\end{table*}

\subsubsection{\textbf{How does Amoeba perform compared to benchmark algorithms?}} Table \ref{table:tor_results} presents the performance of each classifier detecting Tor and V2Ray traffic respectively, as well as the efficacy of adversarial attacks targeting these classifiers. In the absence of adversarial manipulations, the selected classifiers yield almost perfect accuracy and F1 scores (third column) as expected on the \textit{test\_set}, since both anti-censorship systems generate unique statistical patterns during communications. For example, when observed on the TCP layer, Tor traffic mostly consists of packets of (multiples of) 536 bytes, which is the size of an encapsulated onion cell, giving ML classifiers high confidence to detect. V2Ray's TLS-tunneled flows can be differentiated from HTTPS flows, because for HTTPS, once the TLS connection is established, the inner communications are all HTTP requests/responses; while for TLS-tunneled flows, the inner communications may involve a TLS handshake between browser and web server. 
This TLS-in-TLS pattern would not be witnessed in normal browsing traffic without a tunnel, which gives ML classifiers opportunities to learn the discrepancies based on the statistical features.  

On the other hand, the selected white-box adversarial attacks are effective in generating adversarial features of
network flows. It is not surprising that the \gls{CW} attack can reach $\sim$92\%\gls{ASR} on average with $\sim$37\% data and $\sim$18\% time overhead (fourth column). This attack explores misleading perturbations by leveraging the weights and gradients of the censoring classifiers, and iteratively optimizes an adversarial sample for each input (network flow).
However, the practicality of \gls{CW} is questionable in the networking domain, since it requires 1) a complete flow as input; and 2) multiple rounds of queries to the censoring classifiers until a legitimate adversarial flow is found. 

NIDSGAN and \gls{BAP} overcome the second issue by training a neural network to generate perturbations for arbitrary inputs in advance, and in the deployment stage adversarial flows can be generated in one go. NIDSGAN has however limited flexibility, since the length of adversarial flows must be equal to the length of input flows. If the censoring classifiers are able to learn directional features from sensitive flows, simply adding perturbations to each packet without inserting new packets would not change directional features, potentially leading to the failure of NIDSGAN. \gls{BAP} utilizes a dedicated \gls{NN} to learn at which positions in a flow if new packets should be inserted, as an approach to disturb directional features. Based on the results in Table \ref{table:tor_results}, we remark that NIDSGAN and \gls{BAP} have their own merits, but can also be unstable when confronting different \gls{NN} architectures. Since both methods generate perturbations for an entire flow, it would be difficult to learn how the changes of a small number of packets in a flow would impact the final classification results. In contrast, Amoeba is designed to observe the classification result upon every new packet in an adversarial flow, which provides fine-grained information to infer the decision boundary of classifiers.

Table \ref{table:tor_results} reveals that our proposed \textbf{Amoeba reaches $\sim$94\% \gls{ASR} on average against multiple types of classifiers}, 
being capable of exploring the decision boundary of a classifier even if they are not NN-based (and thus offer no gradient information which is required by existing attacks), including \gls{DT}, \gls{RF} and \gls{SVM}/CUMUL. Compared with white-box methods, Amoeba \textbf{follows a much stricter threat model where feature engineering and model architecture are invisible, and it is also more stable against different classifiers}. 
The data overhead of the adversarial flows are in a similar range, between 43.2--64.8\%, except for adversarial samples against \gls{DT}/\gls{RF} on \texttt{Tor Dataset} (where it is lower, yet a comparison with gradient-based methods infeasible) and those for \gls{LSTM} on the \texttt{V2Ray Dataset}. The time overhead of adversarial flows is consistently $<$10.5\%. Appendix \ref{sec:action_analysis} offers an analysis of the actions taken to attack different classifiers.

\begin{figure}
\centering
\vspace{-1em}
\begin{minipage}{.49\textwidth}
  \centering
  \includegraphics[width=\linewidth]{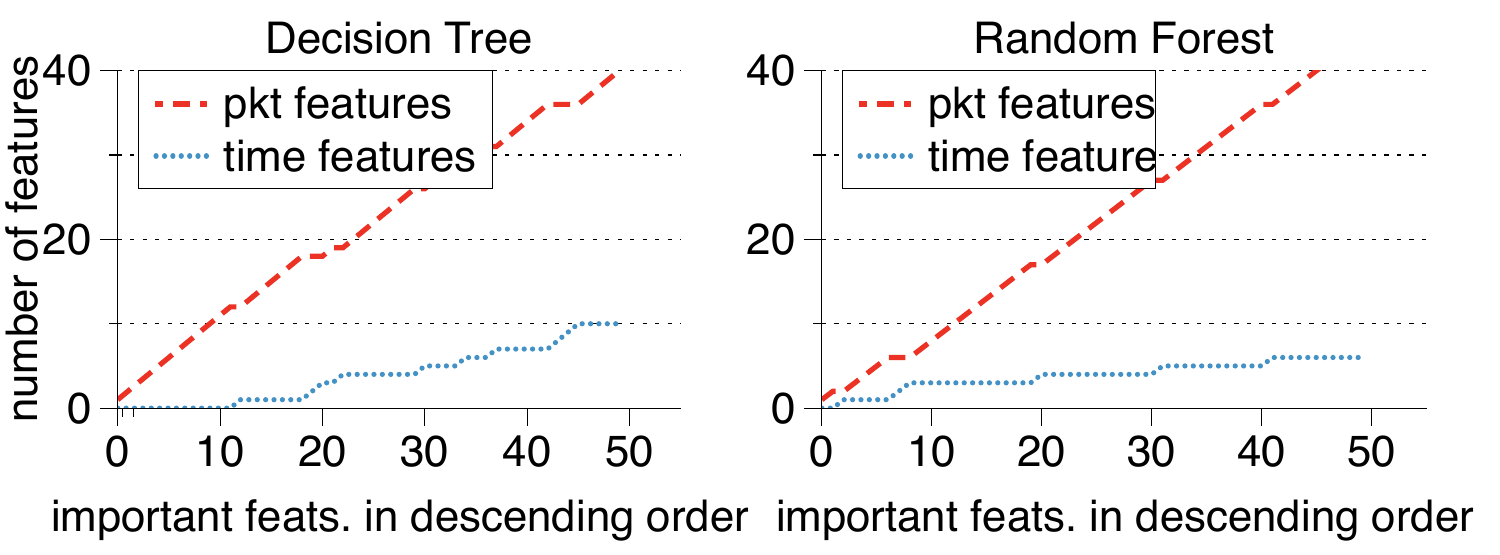}
  \vspace{-2.5em}
  \caption{Difference between packet and timing features  among top-50 important ones used by \gls{DT}/\gls{RF} on  the \texttt{V2Ray dataset}. x-axis arranges features by importance.}
  \label{fig:feature_importance}
\end{minipage}%
\hfill
\begin{minipage}{.49\textwidth}
  \centering
  \includegraphics[width=1.1\linewidth]{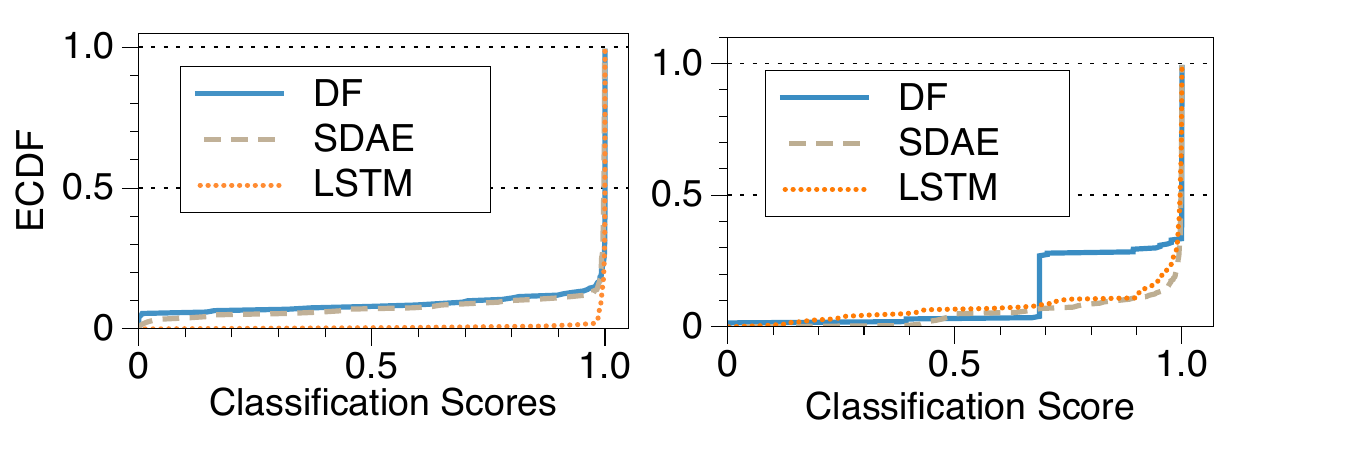}
  \vspace{-2.5em}
  \caption{\gls{ECDF} of classification scores wrt. adversarial flows against different NN-based classifiers. Left plot shows the scores obtained on \texttt{Tor}, right for \texttt{V2Ray}.}
  \label{fig:scores_combined}
  \begin{minipage}{.1cm}
            \vfill
    \end{minipage}
\end{minipage}
\vspace{-2em}
\end{figure}

\subsubsection{\textbf{Is Amoeba sensitive to changes in the network environment?}}
\label{sec:feat_importance}
A shared observation on the results with both datasets is that adversarial flows possess greater data overhead than time overhead. The reason is that censoring classifiers leverage more on packet features than on timing features to make decisions. We visualize important features used by \gls{DT} and \gls{RF} in Figure \ref{fig:feature_importance}, where the x-axis lists top 50 important features in descending order, and the y-axis shows the number of packet and timing features respectively. Observe that packet features in general are overwhelmingly more important than timing features. Practically, network flows may suffer different degree of congestion depending on route and time, while packet/record sizes in a flow are purely determined by the client and the server, thus more reliable for the censoring classifiers. As a result, Amoeba makes more efforts to reshape sizes than timings.

In a more extreme setting where not only network congestion exists but packets are also dropped due to overwhelming volume of traffic in the network, packet retransmission would be needed to tackle data loss. To evaluate the impact of different packet drop rates on the performance of Amoeba, we additionally collect \texttt{Tor Dataset}s multiple times where we enforce packet drop rates for bi-directional traffic between 0\% and 10\%. The same data preprocessing/split convention is followed. We train Amoeba against \gls{DF} on the \textit{attack\_train\_set}s collected under different packet drop rates and then evaluate it on different \textit{test\_set}s. The results are shown in Table \ref{table:different_env}, in which the numbers on the first column represent the packet drop rates under which the training sets are collected, and those on the first row indicate the packet drop rates under which the test sets are collected. The \gls{ASR} [\%] of Amoeba trained and tested under the same environment are shown on the diagonal of the table in bold, and the rest of the numbers indicate the performance difference in \% when cross-evaluating Amoeba in different environments.

\begin{wrapfigure}{l}{0.54\columnwidth}
\vspace{-1em}
      \centering
        \begin{tabular}{c|ccccc}
        \Xhline{2\arrayrulewidth}
        \begin{tabular}[c]{@{}c@{}}Train/Test \\ Pkt Drop Rate\end{tabular} & 0\%  & 2.5\% & 5\%  & 7.5\% & 10\% \\ \hline
        0\%                                                                 & \textbf{87.5} & -8.2  & -8.1 & -6.4  & -7.4 \\
        2.5\%                                                               & -0   & \textbf{88.8}  & -0   & -0.2  & -0   \\
        5\%                                                                 & -2.0 & -1.6  & \textbf{94.2} & -1.2  & -1.2 \\
        7.5\%                                                               & +0.8 & -1.8  & -1.4 & \textbf{94.2}  & -0.4 \\
        10\%                                                                & -1.2 & +0.6  & -1.2 & -0.8  & \textbf{92.0} \\ \Xhline{2\arrayrulewidth}
        \end{tabular}
    \caption{The \gls{ASR} [\%] of Amoeba trained and tested under the same environment are shown on the diagonal of the table in bold, and the rest of the numbers indicate the performance difference in \% when cross-evaluating Amoeba in different environments.}
    \label{table:different_env}
    \vspace{-2em}
\end{wrapfigure}

Amoeba trained with data experiencing 2.5\%--10\% packet drop rates exhibits particularly robust performance when perturbing network flows collected from other environments (2\textsuperscript{nd} to 5\textsuperscript{th} rows). However, if the training set does not incorporate retransmitted packets, it would be more difficult for Amoeba to perturb network flows collected with non-zero packet drop rates (1\textsuperscript{st} row). This is not surprising since the dataset collected with 0\% drop rate is less heterogeneous than that with non-zero drop rates. This set of results reveals that network environment is an important factor when collecting network flows, and if the dataset can reflect the heterogeneity of the network, then Amoeba is less sensitive to changes in the network environment. 

\subsubsection{\textbf{What is the cost of using Amoeba and can that be reduced?}}
\label{sec:reward_masking}
\begin{figure}[h]
\vspace{-0.5em}
    \centering
    \includegraphics[width=\columnwidth]{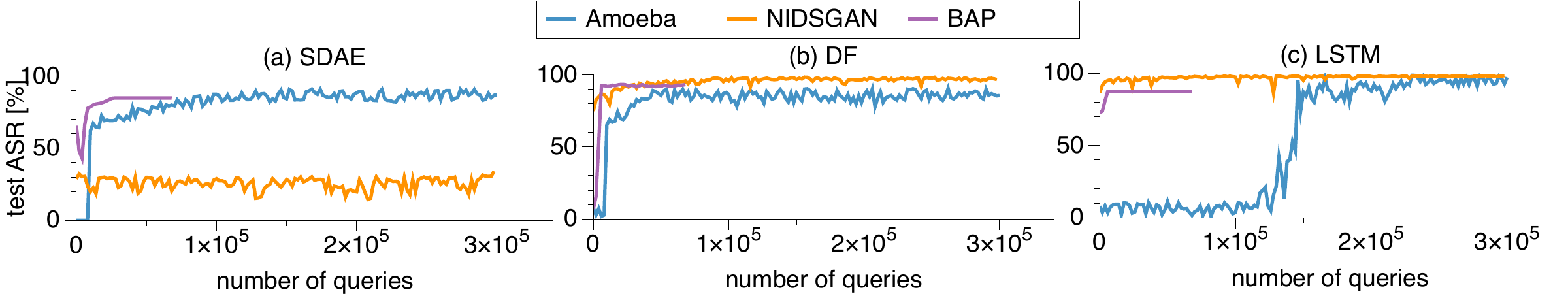}
    \vspace{-2em}
    \caption{Convergence curves of NIDSGAN, BAP and Amoeba attacking three classifiers on \texttt{Tor Dataset}. }
    \label{fig:convergence_compare}
    \vspace{-1em}
\end{figure}

Effectiveness against CUMUL/DT/RF aside (where the benchmarks considered don't work), Amoeba's ASR is higher than or on par with that of \textit{white-box} attacks at the cost of a) higher data overhead, and b) 2 to 10 times more interactions with the censoring classifiers (see Fig.~\ref{fig:convergence_compare}). The reason is two-fold: 1) Amoeba is a black-box algorithm and therefore requires more queries by nature; 2) Given a flow $S$, Amoeba is designed to interact with the classifier at least $|S|$ times and observe associated rewards, whereas \gls{BAP} only needs to interact once in a training epoch. 

\begin{figure}[t]
    \centering
    \includegraphics[width=\columnwidth]{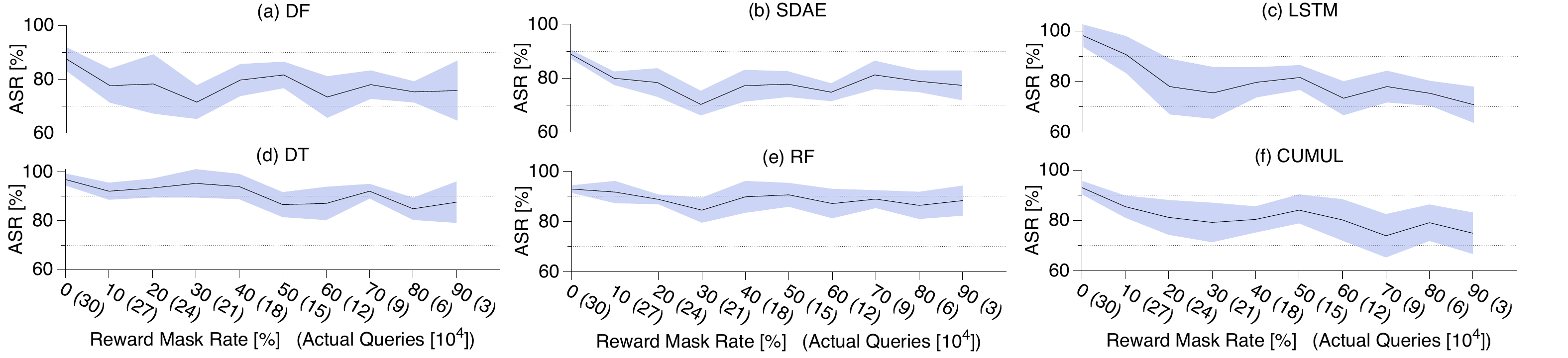}
    \vspace{-2em}
    \caption{Impact of reward mask rate on Amoeba's \gls{ASR}. The reward mask rate is controlled to increase from 0\% to 90\% and the actual number of queries are in the brackets.}
    \label{fig:mask_rate}
    \vspace{-1em}
\end{figure}

\begin{figure}[h]
\vspace{-0.5em}
    \centering
    \includegraphics[width=\columnwidth]{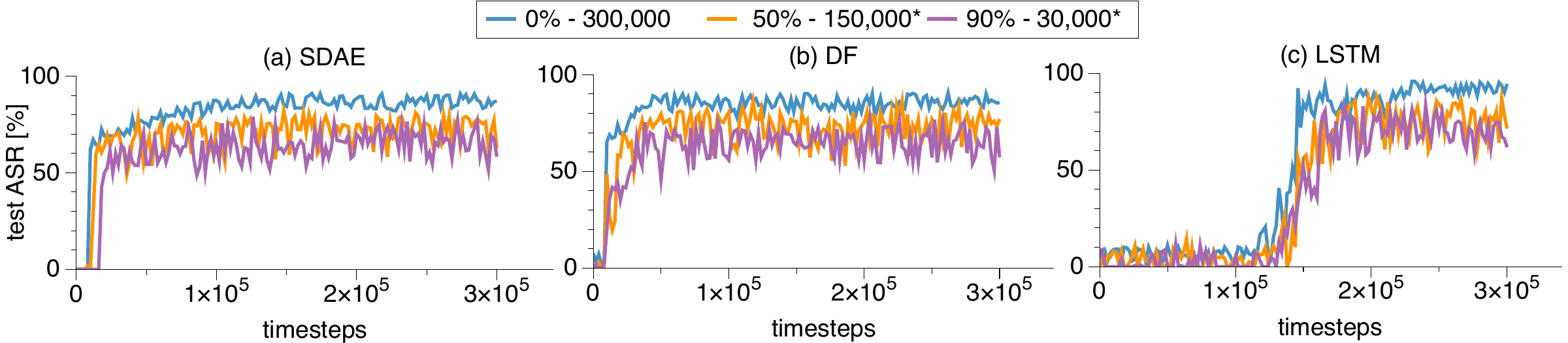}
    \vspace{-2em}
    \caption{Convergence curves of Amoeba attacking three classifiers under different reward mask rates, namely 0\%, 50\% and 90\%. The legend represents the mask rate and the number of queries performed at the end of training. * denotes estimated value given that rewards are randomly dropped. Note that the x-axis represents timesteps instead of the number of queries (for orange and purple curves), because at the timesteps when the rewards are dropped, essentially no query is performed.}
    \label{fig:mask_convergence}
    \vspace{-1em}
\end{figure}

However, in practice it may not be always possible to perform countless queries to censoring classifiers. We therefore attempt to reduce the number of interactions needed by randomly masking the rewards when training Amoeba. In the vanilla version of the training algorithm, Amoeba expects to receive a part of the reward $r_{adv}$ for each subsequence of the generated flows, with 1 denoting good and 0 for sensitive. We mask $r_{adv}$ with a probability $p_{mask}$ from 0\% to 90\% during training, and the masked $r_{adv}$ is considered to be 0.5 instead, representing unknown feedback. Amoeba is trained with 300,000 timesteps, and the actual number of queries would be $300,000 \times (1 - p_{mask})$. Each experiment is repeated 5 times and Fig.~\ref{fig:mask_rate} plots the average \gls{ASR} under each mask rate, with the shaded area representing the $\pm std$ of the results. Amoeba would experience larger variance during training when the reward is randomly masked regardless of the type of censoring classifiers applied. In particular, as the reward mask rate increases from 0\% to 90\%, the \gls{ASR} against \gls{DF}, \gls{SDAE}, \gls{LSTM} and CUMUL drops by 16.5\% on average, whereas the \gls{ASR}s against \gls{DT} and \gls{RF} only drops by 7\% on average. This is because tree-based models utilize flow features for classification \cite{barradas2018effective}, and the absence of the reward for a specific adversarial packet is of lesser consequence, provided that the generated packets adhere to the learned adversarial patterns \textit{in the feature space}. On the contrary, other models using the flow representation in Section \ref{sec:problem_formulation} as inputs can be sensitive to the alterations to \textit{each individual packet}. Lack of accurate rewards at each timestep would challenge Amoeba in learning a reliable approach to generate adversarial flows (see Fig. \ref{fig:mask_convergence}). However, Amoeba is still robust even if the rewards are highly noisy, considering that the number of queries can be reduce by 10$\times$ to 30,000 and the average \gls{ASR} sustained is 79\%.

\subsubsection{\textbf{Are adversarial samples transferable?}}
\label{sec:transfer_tor}

\begin{wrapfigure}{r}{0.7\columnwidth}
\vspace*{-1.5em}
    \centering
    \includegraphics[width=0.7\columnwidth]{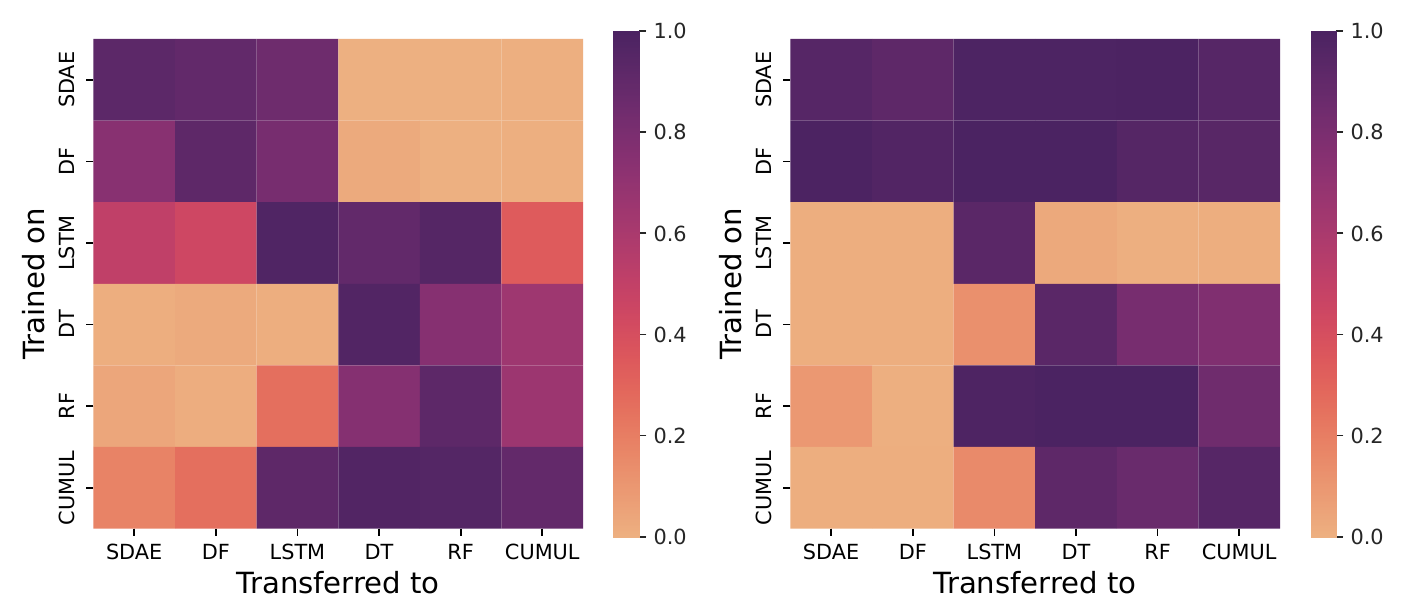}
    \vspace*{-1.5em}
    \caption{Transferability of adversarial flows. Adversarial examples generated by Amoeba against each model on the y-axis and tested on other models on the x-axis. Color of each cell represents \gls{ASR}. 
    The left heatmap is obtained on the \texttt{Tor Dataset} and the right one on the \texttt{V2Ray Dataset}.}
    \label{fig:tor_heatmap}
    \vspace{-1em}
\end{wrapfigure}

Here we investigate whether adversarial flows generated by Amoeba against one classifier can also deceive other models without retraining. To this end, we store all the adversarial samples obtained from each model and feed them to the rest of the classifiers for both datasets. We plot success rates as a heat map in Figure \ref{fig:tor_heatmap}, 
where we find that \textbf{adversarial flows targeting similar architectures are transferable with high success rate}, such as \gls{SDAE} and \gls{DF}, and \gls{DT} and \gls{RF}, meaning that these pairs of classifiers are likely to learn a similar decision boundary. 

The adversarial samples targeting \gls{LSTM} on the \texttt{V2Ray Dataset} are exceptional with only 7.73\% data overhead on average. It is likely that Amoeba uncovers a unique and efficient strategy to attack sequential models on this dataset, but cannot be easily generalized to other censoring classifiers.

\subsubsection{\textbf{How is the quality of adversarial flows?}}
When attacking NN-based classifiers, the architecture and the weights are invisible to Amoeba, but our algorithm can still explore the decision boundary effectively and find qualified adversarial flows. Fig. \ref{fig:scores_combined} plots the \gls{ECDF} of the classification scores with respect to adversarial flows against different NN-based classifiers on both datasets, where the majority of the scores are close to 1 (benign) rather than 0.5. This means that during training, Amoeba does not choose actions randomly in the action space, but \textbf{can understand where the decision boundary lies in the black box and generates adversarial flows just as innocuous traffic}, from the perspective of ML-based classifiers. 

\vspace*{-0.5em}
\subsection{Discussion}
\label{sec:discussion}
\subsubsection{\textbf{Feasibility of deployment}}
Having demonstrated the efficacy of Amoeba, here we discuss the feasibility of deploying our solution in practice. Integrating the algorithm in the transport layer and morphing each packet at line speed is the most ideal way of usage. We run a single-step action inference on a NVIDIA K80 GPU for 10 times and obtain an average inference time of $0.370 \pm 0.001$ ms, which despite small, may be considered non-negligible in the sequence generation process.

\begin{figure}[t]
\vspace{-0.5em}
    \small
    \begin{minipage}{.48\linewidth}
      
    \includegraphics[width=\linewidth]{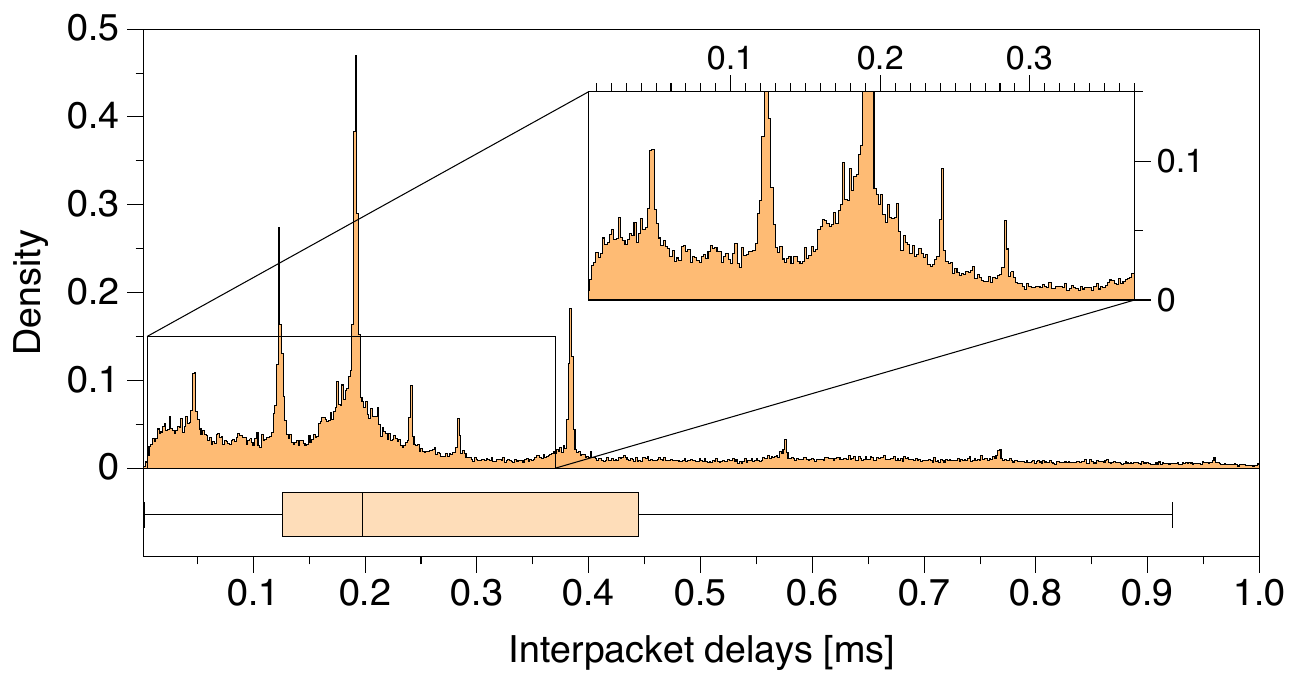}
    \vspace*{-1.5em}
    \caption{The distribution (upper) and box-plot (bottom) of the inter-packet delays between every two consecutive packets in the same network direction.}
    \label{fig:ipd_density}
    \begin{minipage}{.1cm}
            \vfill
    \end{minipage}
    \vfill
    \end{minipage}%
    \hfill
\begin{minipage}{.48\linewidth}
      \centering
        \begin{tabular}{c|cc}
        \Xhline{2\arrayrulewidth}
       \begin{tabular}[c]{@{}c@{}}Censoring\\ Classifier\end{tabular} & \begin{tabular}[c]{@{}c@{}}Data\\ Overhead {[}\%{]}\end{tabular} & \begin{tabular}[c]{@{}c@{}}Time \\ Overhead {[}\%{]}\end{tabular}  \\ \hline
        SDAE                 & 71.22 & 50.02        \\
        DF                   & 76.37 & 63.07       \\
        LSTM                 & 67.99 & 43.44      \\
        CUMUL                & 63.22 & 50.71        \\
        DT                 & 64.53 & 59.68       \\
        RF                 & 60.58 & 38.02      \\\Xhline{2\arrayrulewidth}
        \end{tabular}
    \captionof{table}{Average data overhead and time overhead by embedding tunneled flows into pre-stored adversarial profiles on \texttt{Tor Dataset}. }
\label{table:profile_overhead}
    \end{minipage}
    \vspace{-3em}
\end{figure}

To better understand this, Fig. \ref{fig:ipd_density} shows the density and the box plot of the inter-packet delays between \textit{every two consecutive packets in the same network direction} in our dataset, where 67.5\% of the inter-packet delays are less than 0.37 ms. The time needed for inference would challenge the deployment of the algorithm in an online manner. However, it is still possible to utilize it once Amoeba is well-trained against a censoring classifier. Specifically, we can generate a number of adversarial flow profiles, which only consist of packet sizes and inter-packet delays without real payloads. The profiles would be saved in a database and synchronized with both client and server proxies. During communications, both parties embed actual payload into flows exactly as the flow profiles instruct. If one end has no payload in the buffer but the profile indicates a packet should be sent, then a packet with dummy payload would be transmitted to align with the pre-generated adversarial flow. Although this approach may further increase data overhead, since the flow profiles are not generated based on the current states, it ensures that ML-supported censorship can be successfully circumvented. To illustrate the overhead involved, we store all the adversarial flows (profiles) in the training set of the \texttt{Tor Dataset} that successfully circumvent each classifier, and embed tunneled flows in the test set into the pre-stored, adversarial profiles, as described above. Table \ref{table:profile_overhead} lists the data and time overhead against each censoring classifier respectively. Note that the increase in time overhead is much larger than that in data overhead, by comparing columns 1 and 2 in Table \ref{table:profile_overhead}, since it is common to use multiple adversarial profiles to transmit a single tunneled flow, resulting in extra TCP handshakes to establish connections.
More engineering efforts, such as matching optimal adversarial flow profiles with IP addresses, can be explored for better user experience. To fully achieve online deployment of Amoeba, technical advances, such as designing dedicated hardware that embeds NICs with computational processors \cite{dpu,smartnic}, are needed.

\subsubsection{\textbf{Interactions with Censorship Systems}} Training Amoeba requires hardware advance to accelerate the computation of forward passes. Besides, the training algorithm engages in frequent interactions with the censoring model until a policy is discovered, and in this process, Amoeba may fail to generate adversarial flows, resulting in the blocking of IP addresses or port numbers. Therefore, we assume that attackers can manage a multitude of IP addresses on both sides of the firewall to cope with prompt responses from the censor, as captured by our adversarial model outlined in Section \ref{sec:threat_model}. In a practical scenario, censors tend to block the destination IP addresses or specific port numbers \cite{hoang2021great,wang2017your}, but hardly take actions on the source IP. This  makes sense since the host initiating connections is often behind NAT, and blocking the source IP would prevent a large group of users from accessing the Internet. On the other hand, Amoeba would require more IP addresses outside the firewall. Tools such as 
MassBrowser \cite{nasr2020massbrowser} provide a peer-to-peer tunneling approach to circumvent censorship, facilitated by numerous volunteers establishing proxies in unrestricted regions, and a similar design may be utilized to train Amoeba.

Another critical issue when interacting with censorship systems is that observing rewards is not always straightforward, as the censor would not inform the classification results, but the attackers have to infer decisions instead. Although time-consuming, one viable strategy involves iteratively establishing connections and incrementally generating new packets in each connection. At the point when a connection cannot be built due to IP/port blocking, the attacker discerns that the preceding connection triggered an alarm. Otherwise, the rewards can be perceived as 1 (benign). This method is effective against a censorship system that promptly responds to unwanted traffic. For example, it was observed that HTTPS connections with ESNI would be blocked by the Great Firewall within 1 second after the censor observes a TLS Client Hello \cite{bock2021your}. A more common scenario may be that the rewards are only observable at a certain timestep (after observing the first $n$ packets \cite{wang2015seeing}, after a flow terminates \cite{barradas2018effective}, or for client-to-server packets only \cite{wu2023great}) rather than upon every adversarial packet being generated. This rationale motivated the experiments conducted in Section \ref{sec:reward_masking}, demonstrating that observing only 
$1/10$ of the rewards can still facilitate the progress of Amoeba, albeit with a reduced ASR. The censor, on the other hand, may collect adversarial flows generated by Amoeba, to enrich the dataset of sensitive connections and train the censoring classifier repeatedly. This would nullify the old policy learned by Amoeba and re-training would become necessary. Whether iteratively training the censoring classifier and Amoeba would reach any equilibrium or  one model would outperform the other alternatively is yet to be determined. This problem may align with the SeqGAN framework \cite{yu2017seqgan}, but has not been explored in network traffic generation, which makes it a potential direction for future research.

\subsubsection{\textbf{Ethical Implications}} Although in this work all the data is collected in a controlled environment without real users attempting to evade censorship, certain ethical implications are to be considered, in the sense that the proposed algorithm involves interactions with a censorship system, which may be illegal in restricted regions and may endanger users/attackers. However, given the black-box nature of censorship, interacting with the system is essentially the only way to understand how it works, which is also the methodology followed by prior studies \cite{bock2021your,wu2023great,gfwtor,frolov2019use,ensafi2015examining,bock2019geneva}.

\section{Related Work}
\label{sec:related_work}
\textbf{Censorship techniques}. Internet censorship is carried out in a number of countries in the world, including China, Iran, Russia and India, to block unwanted communications/services. 

IP filtering and DNS poisoning is the most straightforward method to prevent users from establishing connections. 
For example, both DNS resolution and TCP connections to Google Services fail in China, as Google is on the \gls{GFW}'s blacklist \cite{hoang2021great}. 
Besides, \gls{DPI} can inspect application-layer contents for protocol identification. Tor clients use a unique cipher suite during TLS handshakes, which allows the \gls{GFW} to narrow down the suspected targets of Tor connections \cite{frolov2019use}. Active probing involves sending carefully crafted probes to suspicious servers to determine whether they support forbidden protocols, which 
works against Tor, Shadowsocks, Lantern and obfuscated SSH \cite{frolov2019use, beznazwy2020china}. In recent years, ML algorithms (Decision Tree-based, SVM, etc.) were adopted to detect sensitive network flows. \cite{wang2015seeing, barradas2018effective}. {ML-supported censorship may appear similar to \gls{WF} \cite{panchenko2011website,juarez2014critical,panchenko2016website,rimmer2018,sirinam2018deep,sirinam2019triplet}, with the difference that the former targets forbidden network protocols and the latter identifies specific websites. Existing traffic analysis models for \gls{WF} can be easily adopted for network censorship as our results indicate. }

\noindent\textbf{Censorship circumvention approaches.} SkypeMorph \cite{mohajeri2012skypemorph} changes the packet distribution of Tor's traffic to look like connections initiated by Skype. ScrambleSuit \cite{winter2013scramblesuit} and obfs4 \cite{obfs4} add random padding to each packet  to eliminate the fingerprints of fixed-size onion cells in Tor. Tunneling tools embed covert messages into cover protocols, e.g. Meek \cite{fifield2015blocking} tunnels Tor traffic over HTTPS connections. V2ray \cite{v2ray} supports a range of tunnels including HTTP, TLS and Shadowsocks. DeltaShaper \cite{barradas2017deltashaper} transforms covert data into images and transfers them in Skype videocalls. Unfortunately, the aforementioned tools may be vulnerable to ML classifiers \cite{houmansadr2013parrot,wang2015seeing,barradas2018effective}. Protozoa \cite{barradas2020poking} hijacks the WebRTC stack in Chromium and transmits hidden messages through real-time video streaming apps. \textcolor{black}{Geneva \cite{bock2019geneva} and SymTCP \cite{wang2020symtcp} design automated algorithms to discover the vulnerabilities of stateful \gls{DPI} system implemented by censors.}
CDN browsing \cite{zolfaghari2016practical} hosts different web resources on the same set of IP addresses, and provide fake SNI in TLS handshakes to misguide censors.
Decoy routing \cite{nasr2017waterfall} leverages `friendly' Internet autonomous systems which forward messages to the covert destinations. However, these systems are non-standard compliant.

\noindent\textbf{Adversarial attacks against ML classifiers.} The majority of adversarial attacks are confined to computer vision and very few apply to the networking domain. For example, \gls{FGSM} \cite{goodfellow2014explaining} is an effective white-box attack that finds adversarial examples through the gradients of making a wrong prediction. Square attack \cite{andriushchenko2020square} considers the victim classifier to be a black box and randomly adds perturbations to a small patch of the image, until an adversarial example is found. Another strategy of conducting black-box attacks involves a two-stage approach: 1) substituting model training; 2) adversarial sample crafting, which does not directly infer the decision boundary of ML classifiers, but leverages the transferability of samples obtained from the substitute model \cite{papernot2017practical,hang2020ensemble,usama2019black}. However, as evidenced by our results in Section \ref{sec:transfer_tor}, adversarial flows are not always transferable if the true architecture is distinct from the substitute model.

Recent research attempts to use ML to obfuscate traffic features. \gls{GAN} showed ability to generate network flow features that are indistinguishable by ML classifiers \cite{sheffey2019improving, li2019dynamic}. However, only manipulating at feature level is impractical, since mapping features back to a legitimate flow is challenging. iPET \cite{petspaper} and NIDSGAN \cite{zolbayar2022generating} proposes \gls{GAN}-based methods to generate perturbations on network traffic directly as an attempt for deceiving ML classifiers. Apart from adding perturbations to existing packets, \gls{BAP} \cite{nasr2021defeating} learns the optimal position in a flow where to insert dummy packets, disturbing directional features.

\section{Conclusions}
\label{sec:conclusion}

In this paper we introduced Amoeba, an original black-box attack based on adversarial reinforcement learning for circumventing ML-based network traffic censoring classifiers. We demonstrated empirically that Amoeba can shape user flows of arbitrary length over both Tor and V2ray into sequences of packets that have on average 94\% success rates in subverting a broad range of classifiers, and performs stably in different network environments. Amoeba can be trained with considerably noisy rewards and adversarial samples are transferable across similar architectures, proving its robustness and practicality compared with existing attacks. Finally, we provided guidance on how to deploy our solution on real-world systems.

\bibliographystyle{ACM-Reference-Format}
\bibliography{main}

\clearpage
\begin{appendices}
\section{Appendix}

\subsection{Actor \& Critic Optimization}
\label{sec:optimization_details}
We adopt a few optimization tricks in training our agent, including (1) surrogate objective function \cite{schulman2017proximal}; (2) reparameterization trick, and (3) parallel rollout\cite{schulman2017proximal}:

\noindent\textbf{(1) Surrogate Objective:} Directly optimizing Eq. \ref{eq:actor_loss} using a sampled trajectory through multiple steps of gradient ascent may lead to overwhelmingly large, and sometimes worse policy updates. \gls{TRPO} \cite{schulman2015trust} and \gls{PPO} \cite{schulman2017proximal} propose to use a surrogate objective function which theoretically guarantees policy improvement over stochastic gradient ascent:
\[
\max_{\theta} \mathbb{E}_t\left[\frac{\pi_\theta\left(a_t \mid s_t\right)}{\pi_{\theta_{\text {old }}}\left(a_t \mid s_t\right)} A(s_{t}, a_{t})\right],
\]
in which $\theta_{old}$ represents the parameters of an older version of the actor network in stochastic optimization. The surrogate objective function intuitively encourages the actions with positive advantages $A(a_{t}, s_{t}) > 0$ and discourages the opposite. We follow the \gls{PPO} design to clip the ratio $I_{t}(\theta) = \frac{\pi_\theta\left(a_t \mid s_t\right)}{\pi_{\theta_{old}}(a_{t} | s_{t})}$ (avoiding excessive update steps), and add an entropy term to encourage exploration in the action space in the final version of the objective function:
\begin{align}
    \begin{split}
\max_{\theta} \mathbb{E}_t [ \min I_{t}(\theta) A(s_{t}, a_{a}), &clip(I_{t}(\theta), 1-\epsilon, 1+\epsilon) A(s_{t}, a_{a})) ] \\ &+ H_{a_{t} \sim \pi_{\theta}}(a_{t}) 
    \end{split}
\end{align}
\noindent\textbf{(2) Reparametrization trick:}  $\pi_{\theta}(\cdot)$ should approximate the distribution of actions given states but a simple \gls{MLP} network only generates deterministic outputs. To overcome this issue, we assume that all the actions are sampled from a Gaussian distribution, and make $\pi_{\theta}$ generate the mean and the standard deviation of actions given states $\bar{a}_{t}, \sigma = \pi_{\theta}(s_{t})$, as shown in Figure \ref{fig:addpg}. An action then can be sampled by:
\[
a_{t} = \bar{a}_{t} + \epsilon \sigma, \epsilon \sim \mathcal{N}(0, 1).
\]
The trick ensures the actor network is differentiable, as well as generating probabilistic outputs during training.

\noindent\textbf{(3) Parallel Rollout:} In order to speed up model convergence, \gls{PPO} \cite{schulman2017proximal} proposes to train the agent with parallel environments ($N$ in total) where trajectories would be sampled from each environment independently with a fixed timestep $T$, resulting in $N \times T$ observations each time (Algorithm \ref{alg:amoeba_trianing} line 4). The advantage at every timestep is estimated via generalized advantage estimation \cite{schulman2015high}:
\[
A_{t} \approx \sum_{l=0}^{\infty} (\gamma \lambda)^{l} [r_{t+l} + \gamma V(s_{t+l+1}) - V(s_{t+l})],
\]
in which $\gamma$ is the discount factor and $\lambda$ balances the bias and the variance of advantage estimation. We set $\gamma = 0.99$ and $\lambda = 0.95$. If one trajectory terminates before step $T$, the environment starts to generate a new one until reaching $T$ steps and if the trajectory does not terminate after $T$, the advantage can still be estimated by $A_{T} \approx r_{T} + \gamma V(T+1) - V(T)$. $N \times T$ observations along with the actions, the returns and the estimated advantages are then even split into $K$ mini-batches for stochastic optimization (Alg. \ref{alg:amoeba_trianing} line 11-13). The full training algorithm is detailed in Algorithm \ref{alg:amoeba_trianing}.

\begin{wrapfigure}[16]{L}{0.6\textwidth}
\vspace{-2.5em}
\begin{minipage}{0.6\textwidth}
\begin{algorithm}[H]
\small
  \caption{The training algorithm for StateEncoder}
  \label{alg:stateencoder_trianing}
  \begin{algorithmic}[1]
    \Inputs{$dataset := \{S_{1},...,S_{n}\};\;\; S_{i} := [s_{i, 1}, ..., s_{i, T}]$}
    \Initialize{Denote $\mathcal{E}(\cdot)$ a two-layer \gls{GRU} StateEncoder and $\mathcal{D}(\cdot)$ as the decoder with the same architecture. $\mathcal{E}$ and $\mathcal{D}$ initialized via Xavier initialization~\cite{glorot2010understanding}. }
    \While {model has not converged}
    \For{$S_{i}$ sampled from $dataset$}
    \State {$S_{i} \gets S_{i}[:t], t \sim \mathcal{U}(1, T)$}
    \State {$\hat{S}_{i} \gets \mathcal{D}(\mathcal{E}(S_{i})) $}
    \State {$\mathcal{L} \gets MAE(S_{i}, \hat{S}^{i})$}
    \State {$
        \mathcal{E}, \mathcal{D} \gets Adam(\mathcal{L}, \mathcal{E}, \mathcal{D})$}
    \EndFor
    \EndWhile
    \State {\Return {$\mathcal{E}$}}
    \end{algorithmic}
\end{algorithm}
\end{minipage}
\end{wrapfigure}

\subsection{StateEncoder}
\label{sec:stateencoder}
As mentioned in Section \ref{sec:rl_summary}, the state at timestep $t$ is the history of the observations and the actions, meaning that the length of the state would vary as $t$ increases. However, if the agent is built with non-recurrent neural networks, such as MLP or CNN, it requires inputs of fixed size. To overcome this problem, we design StateEncoder, a two-layer, pre-trained \gls{GRU} that encodes an arbitrary long network flow to a fixed-size hidden representation. As shown in Figure \ref{fig:state_encoder}, to ensure that StateEncoder $\mathcal{E}$ can properly encode network flows without nontrivial information loss, we train $\mathcal{E}$ as the encoder part of a Seq2Seq Autoencoder, in which StateDecoder $\mathcal{D}$ shares the same architecture with $\mathcal{E}$. Consider a network flow $S = [s_{1}, ..., s_{T}]$ with $T$ packets. $\mathcal{E}$ aims to map $S$ as an representation in the $H$-dimensional hyperspace, $r_{S} = \mathcal{E}(S) \in \mathbb{R}^{H}$, and $\mathcal{D}$ aims to reconstruct the flow from the hidden representation, $\hat{S} = \mathcal{D}(r_{S}) \in \mathbb{R}^{T \times 2}$. We train the Seq2Seq Autoencoder with a \gls{MSE} loss function, i.e.,
\[
L(S, \hat{S}) = \frac{1}{T}\sum^{T}_{t=1} (s_{t} - \hat{s}_{t})^{2},
\]
by the Adam algorithm \cite{kingma2014adam}. The only connection between $\mathcal{E}$ and $\mathcal{D}$ is the hidden representations. Therefore, $\mathcal{E}$ has to encode the input as intact as possible, to ensure that $\mathcal{D}$ can properly reconstruct. Since the StateEncoder is designed to encode heterogeneous network flows effectively, it should be fed with as many distinct flow as possible during training, with a view to acquiring strong generalization abilities. To this end, we create a synthetic, normalized dataset with maximal variability in both packet sizes and time delays, where each packet size $p_{i}$ and inter-packet delay $\phi_{i}$ in the flows are created~via:
\[
p_{i} \sim \mathcal{U}(-1, 1); \;\; \phi_{i} \sim \mathcal{U}(0, 1), i \in [1, T],
\]
with $\phi_{1} = 0$. We assume that all the packet sizes and delays are 0-1 normalized in this dataset. $p_{i}$ is sampled from $\mathcal{U}(-1, 1)$ because the flow is bidirectional. 
We create a training set with 12,000 flows and a test set with 3,000 flows. Since a reward is given at each timestep in an episode, the StateEncoder must be able to encode a sequence with a arbitrary length. Thus, the sequence length of each mini-batch during training is randomly sampled from $[1, T]$ to avoid that State\-Encoder can only encode fixed-size flows. The complete training algorithm is detailed in Algorithm \ref{alg:stateencoder_trianing}. After training, we only preserve $\mathcal{E}$ to encode states for the adversarial actor and critic. 

\subsection{Performance of StateEncoder}
\label{sec:perf_stateencoder}

\begin{figure}[t]
\centering
\begin{minipage}{.49\textwidth}
    \centering
    \includegraphics[width=\linewidth]{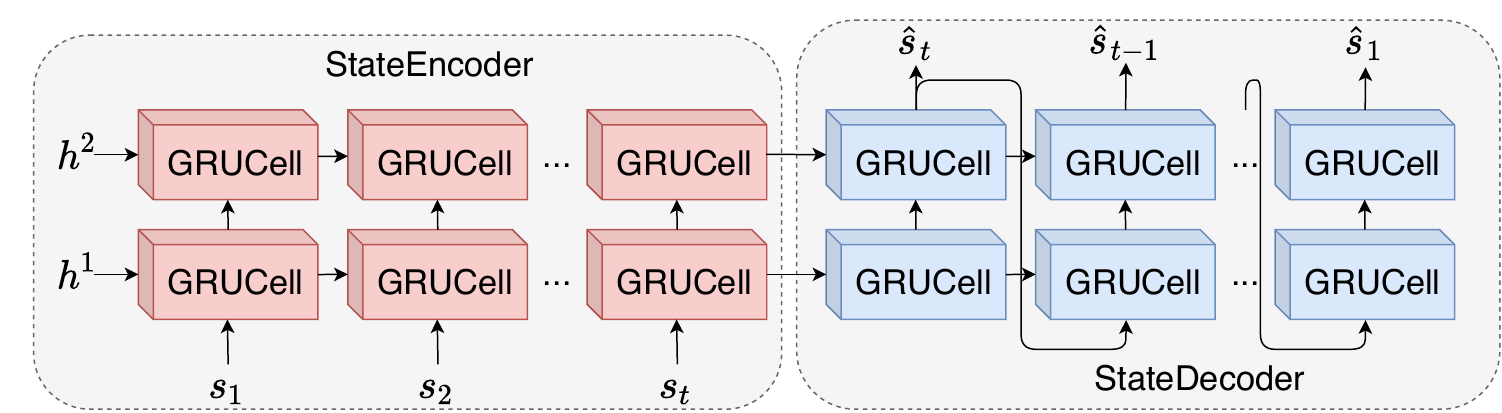}
    \caption{StateEncoder and associated decoder for sequence-to-sequence training. During training, State\-Encoder maps an arbitrary long network flow to a fixed-size hidden representation, which is passed to StateDecoder for reconstruction.}
    \label{fig:state_encoder}
\end{minipage}%
\hfill
\begin{minipage}{.45\textwidth}
    \centering
    \includegraphics[width=\linewidth]{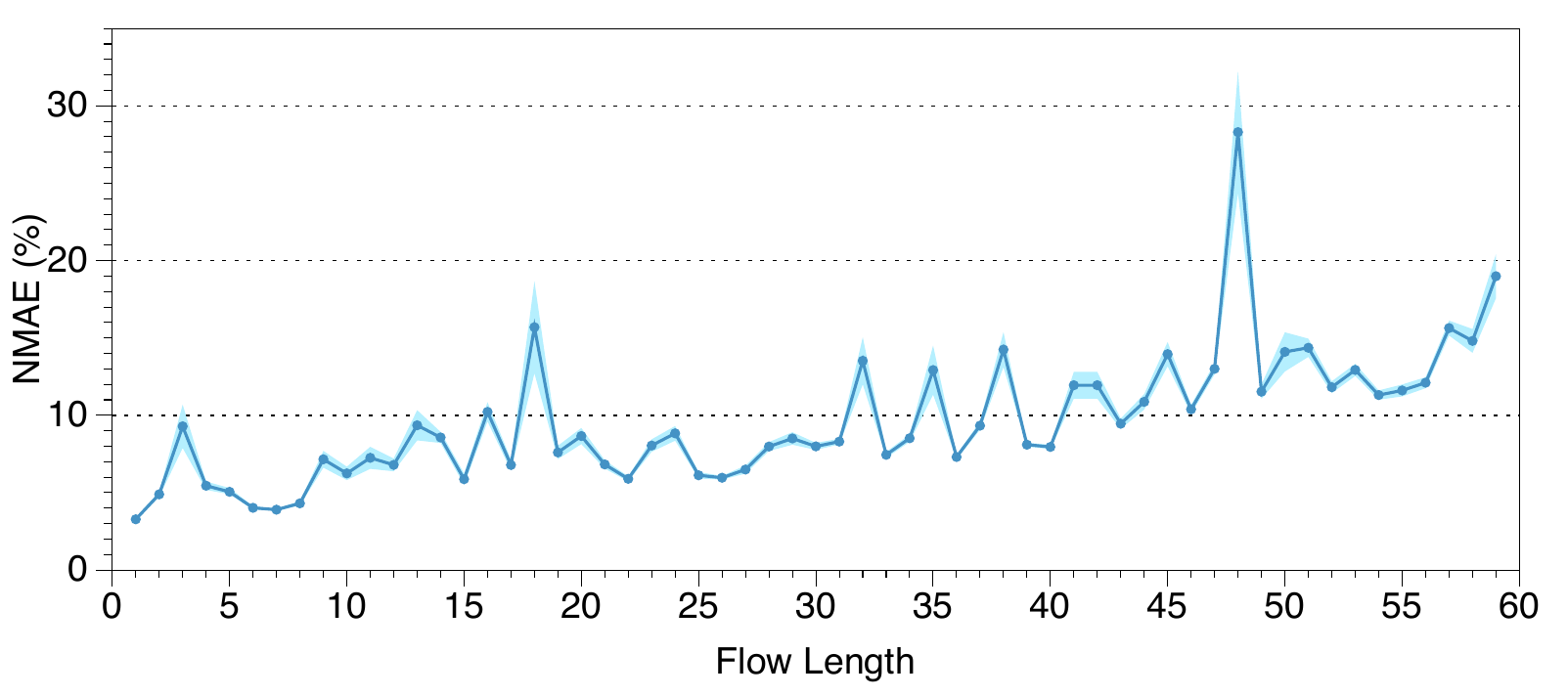}
    \vspace*{-1em}
    \caption{Normalized reconstruction errors (with error bars) of StateEncoder + StateDecoder.}
    \label{fig:recon_tor}
  \begin{minipage}{.1cm}
            \vfill
    \end{minipage}
\end{minipage}
\vspace{-1em}
\end{figure}

The performance of our StateEncoder impacts the adversarial actor in terms of understanding and interpreting the actual state at each timestep. There is no simple method to evaluate StateEncoder alone, since the hidden representations are in high-dimensional space and information loss during encoding is intractable. Nevertheless, we can obtain an upper bound of the information loss by examining the \gls{NMAE} of the Seq2Seq model (consisting of StateEncoder and StateDecoder):
\[
\text{NMAE}(S, \hat{S}) = \frac{1}{T \times N}\sum^{N}_{n=1}\sum^{T}_{t=1} \frac{|s^{n}_{t} - \hat{s}^{n}_{t}|}{s^{n}_{t}}.
\]
$s^{n}_{t}$ is the packet $t$ in flow $n$ and $\hat{s}^{n}_{t}$ the reconstructed packet.
We show the \gls{NMAE} of the Seq2Seq model composed by StateEncoder and StateDecoder in Figure \ref{fig:recon_tor}, which helps us understand to what extent the encoded hidden representations can sustain the information of the original flows. It can be seen that the \gls{NMAE} of flow reconstruction would increase as the flow length increases, although this is not obvious when the flows have less than 40 packets and the average NMAE in $[1, 40]$ is around 9\%. When the flow length is longer than 40, the \gls{NMAE} gradually increases from 9\% to 19\% with an outlier of 28.95\% when the flow length equals 48. An intuitive explanation of the \gls{NMAE} in our case is that, for example, when a flow has 60 packets, each packet size $p$ is encoded as a value between $p \pm 0.19p$ in the hidden representation. Although not perfect, these experiments demonstrate that this level of precision is actually adequate for Amoeba to learn an effective policy. Note that 90.5\% of Tor flows in the dataset have less than 60 packets. To ensure that long flows can be encoded properly in practice, an engineering solution is splitting long flows before a pre-set threshold or using deeper networks to encode flows.

\begin{table}[h]
\begin{tabular}{ccc}
\Xhline{2\arrayrulewidth}
Hyperparameter                                                      & Search Space       & Value               \\ \hline
optimizer                                                           & Adam, SGD, RMSProp  & Adam                \\
learning rate                                                       & {[}0.0001, 0.01{]} & 0.0005              \\
$\lambda_{split}$                                                              & {[}0.01, 0.1{]}    & 0.05                \\
$\lambda_{time}$                                                                & {[}0.1, 2{]}       & 0.2                 \\
$\lambda_{data}$ for \texttt{Tor}                                                               & {[}0.1, 5{]}       & 0.2                 \\
$\lambda_{data}$ for \texttt{V2ray}                                                               & {[}0.1, 5{]}       & 2                 \\
\begin{tabular}[c]{@{}c@{}}Actor/Critic\\ layer number\end{tabular} & {[}2, 5{]}         & 4                   \\
\begin{tabular}[c]{@{}c@{}}Actor/Critic\\ layer dim\end{tabular}    & {[}32, 1024{]}     & \begin{tabular}[c]{@{}c@{}}256$\rightarrow$ 64$\rightarrow$ 32\\ $\rightarrow$output\end{tabular} \\
\begin{tabular}[c]{@{}c@{}}StateEncoder\\ architecture\end{tabular} & {[}GRU, LSTM{]}    & GRU                 \\
StateEncoder dim                                                    & {[}256, 1024{]}    & 512                \\
StateEncoder layer                                                  & {[}1, 4{]}         & 2                   \\ \Xhline{2\arrayrulewidth}
\end{tabular}
\caption{Hyperparameter selection for Amoeba.}
\label{table:hyperparameter}
\vspace{-3em}
\end{table}

\subsection{Hyperparameter Selection}
\label{sec:hyper_selectioin}
Amoeba is a complex model with a range of hyperparameters and it would be difficult to conduct exhaustive search in the full hyperparameter space. To select hyperparameters for Amoeba, We first choose the search space by our experience and build the model in a block-by-block fashion. StateEncoder requires pretraining and therefore the associated hyperparameters are decided initially, followed by the architecture of actor and critic. $\lambda_{data}$, $\lambda_{time}$ and $\lambda_{split}$ plays an important role in the reward function and largely determines the final \gls{ASR} and overhead rates. We notice that Amoeba is not sensitive to $\lambda_{time}$ but the results may vary greatly given different $\lambda_{data}$. Since \texttt{Tor Dataset} and \texttt{V2ray Dataset} have different largest transmission units (TCP segement and TLS record), the optimal $\lambda_{data}$ are 0.2 and 2 respectively. $\lambda_{split}$ determines how frequently the packet should be truncated. Our experimental results indicate that when $\lambda_{split} > 0.1$, Amoeba would tend not to truncate packets at all. If directional features need to be disturbed, $\lambda_{split}$ should be set smaller than $0.1$. We set $\lambda_{split} = 0.05$ eventually so that packet truncation would occur but is not so frequently that exceeds the capability of StateEncoder. We choose the set of hyperparameters in Table \ref{table:hyperparameter} which is good enough to provide high \gls{ASR} and acceptable overhead rates, but there may exist better selections.  

\subsection{Action Analysis}
\label{sec:action_analysis}
\begin{figure}[h]
\vspace{-1.5em}
    \centering
    \includegraphics[width=\columnwidth]{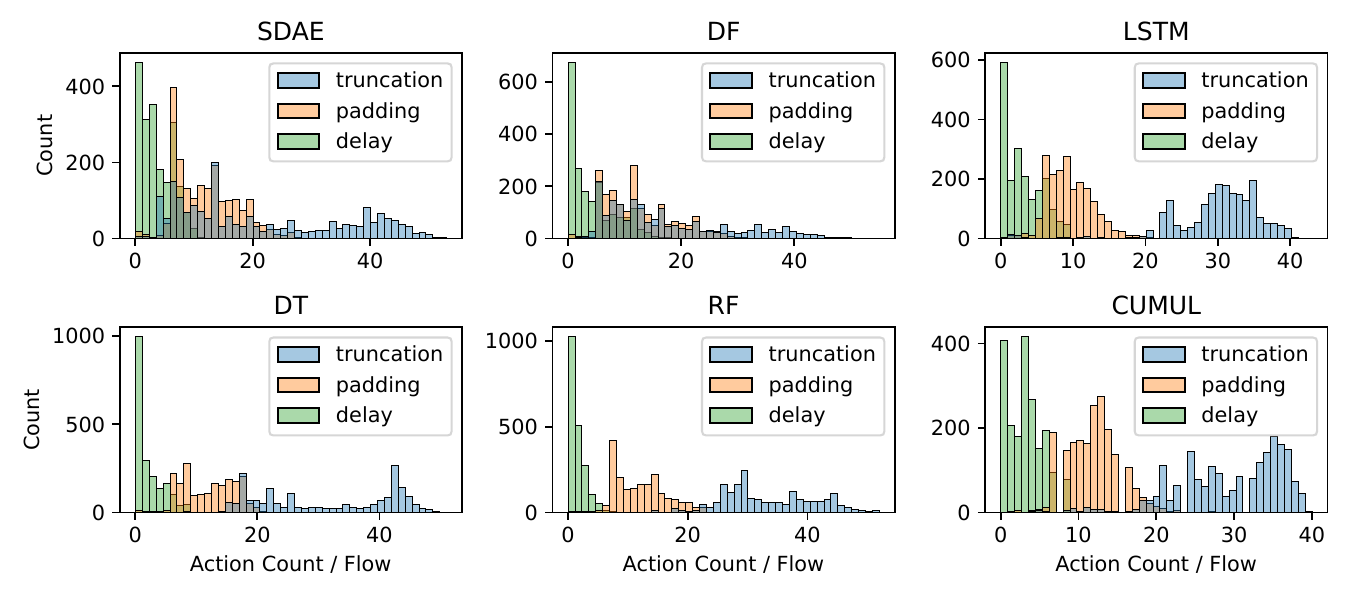}
    \vspace{-2.85em}
    \caption{Histograms of the actions taken per flow (2000 flows in total) to generate adversarial samples against each classifier on \texttt{Tor Dataset}.}
    \label{fig:actions}
    \vspace{-0.5em}
\end{figure}
The time overhead of adversarial flows generated by Amoeba is consistently and significantly lower than the data overhead, as shown in Table \ref{table:tor_results}. Here, we scrutinize the actions selected by Amoeba more closely, namely, truncation, padding, and adding delay. Fig. \ref{fig:actions} presents histograms of the number of actions taken per flow (2000 flows in total) to craft adversarial samples against each classifier on the \texttt{Tor Dataset}. The average length of the tunneled flows prior to obfuscation is 24.5 packets. It is obvious that when generating adversarial flows, adding delay is the least favored action, irrespective of the backend censoring classifiers, yielding less than 8 instances of added delays for the majority of the adversarial flows. In comparison, truncation is commonly employed, especially when attacking \gls{LSTM}, \gls{DT}, \gls{RF} and CUMUL. Its usage is roughly twice as often as the number of padding instances, which effectively alters the directional features in the original, sensitive traffic.

\end{appendices}

\end{document}